\documentclass[12pt,a4paper]{article}

\usepackage{tikz}
\usetikzlibrary{arrows.meta}

\usepackage[
  margin=0.75in,
  headsep=10pt    
]{geometry}

\usepackage{graphicx}
\usepackage{epstopdf}   
\usepackage{amsmath}
\usepackage{amssymb}
\usepackage{breqn}      
\usepackage{makecell}
\usepackage{multirow}
\usepackage{booktabs}
\usepackage{siunitx}
\usepackage{dcolumn}
\usepackage{bm}
\usepackage{xcolor}
\usepackage{textcomp}
\usepackage{gensymb}
\usepackage{textcomp}
\usepackage{subcaption}
\usepackage{lineno}
\usepackage{etoolbox}
\usepackage{authblk}
\usepackage{setspace}
\usepackage{float}

\usepackage{times}      
\usepackage[utf8]{inputenc}
\usepackage[T1]{fontenc}
\usepackage{soul} 
\graphicspath{ {./figures/} }

\usepackage[sort&compress]{natbib}
\usepackage[hidelinks]{hyperref}

\newcommand{\edt}[1]{{\color{black}#1}}  
\newcommand{\dev}[1]{{\color{black}#1}}  

\title{Dynamical Characteristics of the Body-Caudal Fin Joint of a Carangiform Swimmer and its Influence on Hydrodynamics}

\author[1,2]{Dev Pradeepkumar Nayak}
\author[1,$^{\dag}$]{Muhammad Saif Ullah Khalid}
\author[2]{Ali Tarokh}

\affil[1]{Nature-Inspired Engineering Research Lab (NIERL), Department of Mechanical and Mechatronics Engineering, Lakehead University, Thunder Bay, P7B 5E1, ON, Canada}
\affil[2]{Department of Mechanical and Mechatronics Engineering, Lakehead University, Thunder Bay, P7B 5E1, ON, Canada}

\affil[$\dag$]{\textit{Corresponding author:} mkhalid7@lakeheadu.ca}

\begin{document}
\maketitle


\begin{abstract}
The hydrodynamics of fish swimming depend on the interaction between the undulation of the body and the flapping of the caudal fin. This study develops a computational framework of a Jackfish-inspired swimmer with an independently mounted caudal fin that pitches passively under fluid forces and a nonlinear torsional spring. The fin synchronizes with the body when damping and stiffness parameters are tuned correctly, producing passive pitching that closely resembles to the displacement of the actively pitching tail. At Re = $3000$, synchronized passive pitching generates coherent hairpin and ring vortices that reinforce streamwise momentum and contribute to thrust, whereas larger phase differences lead to wake spread in lateral direction and drag-dominated behavior. These results reveal that nonlinear peduncle mechanics naturally regulate amplitude, phase, and recoil, offering a biologically inspired pathway toward underwater robotic design using passive kinematics. 
\end{abstract}

\newpage
\section{Introduction}
Nature \edt{shapes} evolution of aquatic organisms through continuous selection pressures, leading to remarkable efficiency in their movement and control. The marine environment, with its demanding hydrodynamic constraints, \edt{drives} species to develop locomotion strategies that allow them to capture prey, evade predators, and maintain stability in unsteady flow. Central to these strategies is the capacity to generate and manipulate vortical structures that improve thrust and maneuverability. Among the \edt{morpholigical} features that support this capability, the caudal fin plays a defining role in swimming performance \edt{for swimmers belonging to the classes of carangiform, subcarangiform, and thunniform, and ostraciiform. Previously,} a large body of work examined the vortex interactions between the body of the swimmer and the caudal fin, highlighting their collective influence on propulsive efficiency and flow control \cite{liu2017computational, khalid2021larger, gao2018independent, lu2025nonlinear, kumar2025mechanical, triantafyllou1995efficient}.

Understanding these mechanisms is increasingly important for the development of autonomous underwater vehicles and bio-inspired robotic swimmers \cite{amal2024bioinspiration, costa2017computational}. Advances in computational modeling also \edt{expands} our ability to probe complex hydrodynamic and behavioral processes. For instance, the integration of deep recurrent Q-networks with immersed-boundary–lattice–Boltzmann methods \edt{yielded} new insight into optimizing swimming behaviors such as Kármán gaiting and prey capture \cite{zhu2021numerical}. These computational developments complement long-standing observations that the caudal fin substantially influences maneuverability and the ability of the swimmer to regulate body–wake interactions \cite{borazjani2010role, liu2017computational, khalid2021larger}. Multiple studies consistently showed that the geometry of the peduncle and caudal fin is closely tied to  the performance in locomotion across species \cite{lauder2000function, hang2022active, khalid2021larger, wang2020tuna, zhang2020specialization}.

In carangiform swimmers and those belonging to the classes mentioned above, the caudal fin is the primary contributor to thrust. Their kinematics reveal that the undulation of the posterior body and fin can be interpreted as a coordinated heaving–pitching mechanism \cite{Akhtar2007Hydrodynamics}. This concept links to the production of thrust since the classical works of Knoller and Betz \cite{knoller1909gesetzedes, betz1912beitrag}. The \edt{beating tail} generates vortical structures that strongly correlate with \edt{thrust production and} propulsive efficiency \cite{triantafyllou1995efficient}. Species relying heavily on posterior-body propulsion invest considerable muscular and elastic energy to drive the caudal fin, and evolutionary adaptation \edt{has} a robust peduncle\edt{, which is} capable of transmitting this energy efficiently \cite{lighthill1970aquatic}. This \edt{specialized anatomical feature} supports the characteristic amplitude envelope of carangiform swimmers, with the largest lateral displacements occurring near the peduncle and fin.

Both experimental and numerical studies \edt{demonstrates} that the passive foils serve as effective analogues for freely swimming fish and their ability to self-propel through fluid–structure interaction \cite{lauder2012passive, feilich2015passive, kim2018effect}. Lauder et al.~\cite{lauder2012passive} demonstrated that even when stiffness \edt{was} held constant, the trailing-edge geometry of a flexible foil \edt{could} create substantial variation in swimming speed, emphasizing the impact of morphological differences among species. Their results indicated that targeted stiffness distribution or variable stiffness actuation \edt{might} lead to further gains in thrust or propulsive efficiency. Complementary studies, such as those by Liu et al.~\cite{liu2019image}, used coupled structural and fluid solvers to determine the material properties and deformation patterns that \edt{governed} the generation of thrust in real fins. \edt{According to} Hang et al.~\cite{hang2022active}, the joint connecting the body to the caudal fin undergoes pronounced flexion near the posterior region, and that passive bending can yield favorable outcomes for swimming economy despite trade-offs in \edt{the} maximum speed. \edt{Moreover, some other} contributions expanded the understanding of fins as multi-functional control surfaces. Fish and Lauder \cite{fish2017control} showed that fins \edt{played} critical roles not only in thrust but also in maneuverability and stability, illustrating the interplay between active muscle-driven kinematics and passive structural response. \edt{Besides,} Behbahani et al.~\cite{behbahani2016design} introduced a passive feathering joint for pectoral fins that reduced drag during the recovery stroke while maintaining desired kinematics during the power stroke.

A wide range of \edt{robotics-related} studies incorporated passive or variably stiff joints to reproduce fish-like behaviors. \edt{For instance,} Qiu et al.~\cite{qiu2022locomotion} developed a tendon-driven robotic fish with an active tail and a passively deflecting caudal fin with variable stiffness, which resulted in tighter turning radii and faster swimming. Chen et al.~\cite{chen2020exploration} proposed a compliant joint with torsional springs for multi-segment robotic swimmers, while Wang et al.~\cite{wang2025design} introduced a modular variable-stiffness passive joint that improved swimming performance through adaptive stiffness changes. \edt{Furthermore,} Lu et al.~\cite{lu2024effect} further demonstrated a hybrid active and passive Fin Ray–based pectoral fin capable of achieving substantially higher thrust compared to traditional passive configurations.

Across these studies, a recurring observation is that \edt{appropriately designed} passive joints between the body and the fin, whether pectoral or caudal, \edt{may} support favorable swimming characteristics. \edt{It is important to note that these previous studies focused only on determining the dynamical characteristics of the swimmers without explaining the governing fluid-structure interactions based physical mechanisms for the swimmers, and these important underlying aspects remained undetermined and} insufficiently understood. \edt{Contextually,} the connection between the peduncle and caudal fin \edt{may be} modeled using nonlinear stiffness \edt{elements, where} such a formulation allows the fin to undergo amplified displacement near mid-stroke and return toward its neutral position through a recoil at larger deflections. This nonlinear behavior supports synchronized locomotion and \edt{was} reported in several recent works \cite{lu2024effect, lu2025nonlinear, bergmann2014effect}. However, the detailed vortex–body interactions that arise in such systems were \edt{unresolved. Our present study aims towards these specific aspects, which establishes the primary novelty of this work.}

The present study examines the hydrodynamic performance of a \edt{real Jack fish-like} swimmer \edt{composed} of an undulating trunk \edt{and a caudal fin, which performs prescribed heaving and flow-induced pitching. This work extends} the concept proposed by Nayak et al.~\cite{nayak2025comparative} and \edt{Gao et al.} \cite{gao2018independent} \edt{from a two-dimensional swimmer to a more realistic three-dimensional ($\mbox{3D}$) one}. \edt{Very recently, Lu et al. \cite{lu2025nonlinear} and \cite{lu2026dynamic} also introduced a similar concept by only considering the tail, while modeling it as a simple two-dimensional plate. It is important to highlight that their work did not incorporate a swimmer's body in the system, which left important aspects of nonlinear body-tail or vortex-tail interactions entirely unexplored for determining the hydrodynamic performance of relevant biological or bio-inspired swimmers. Our present study also addresses this important gap and provides insights about these highly nonlinear interactions between different physiological parts of a more realistic carangiform swimmer.} 

\edt{Here,} we employ an fluid-structure interaction ($\mbox{FSI}$)–based \edt{computational} framework in \texttt{OpenFOAM} v2312 to solve the Navier-Stokes equations \edt{for unsteady incompressible flows around the three-dimensional swimmer} at a Reynolds number ($\mathrm{Re}$) of $3000$. The specific objectives of this work \edt{include}: (i) \edt{investigating} the performance of a passively pitching caudal fin with a linear and nonlinear torsional spring and damper, \edt{where it also performs prescribed heaving to follow the undulatory kinematic profile of the swimmer}, (ii) \edt{examining} how the damping ratio influences the synchronization between the passive pitching response \edt{of the caudal fin} and undulation of the body, (iii) \edt{comparing} the hydrodynamic performance of the passively pitching configuration with that of an actively pitching caudal fin, and (iv) \edt{analyzing} the role of passive pitching \edt{of the swimmer's} tail in shaping the vortex dynamics that \edt{governs} the fluid–fin \edt{interactions}. By addressing these questions, this study \edt{reinforces foundations} for designing \edt{next-generation} nature-inspired \edt{underwater} robotic platforms capable of improved propulsion through \edt{effective interactions between the swimmer and the surrounding water}.
\section{Computational Methodology}

\subsection{Geometry and kinematics of \edt{the swimmer}}

\begin{figure}[ht!]
    \centering
    \includegraphics[width=1.0\linewidth]{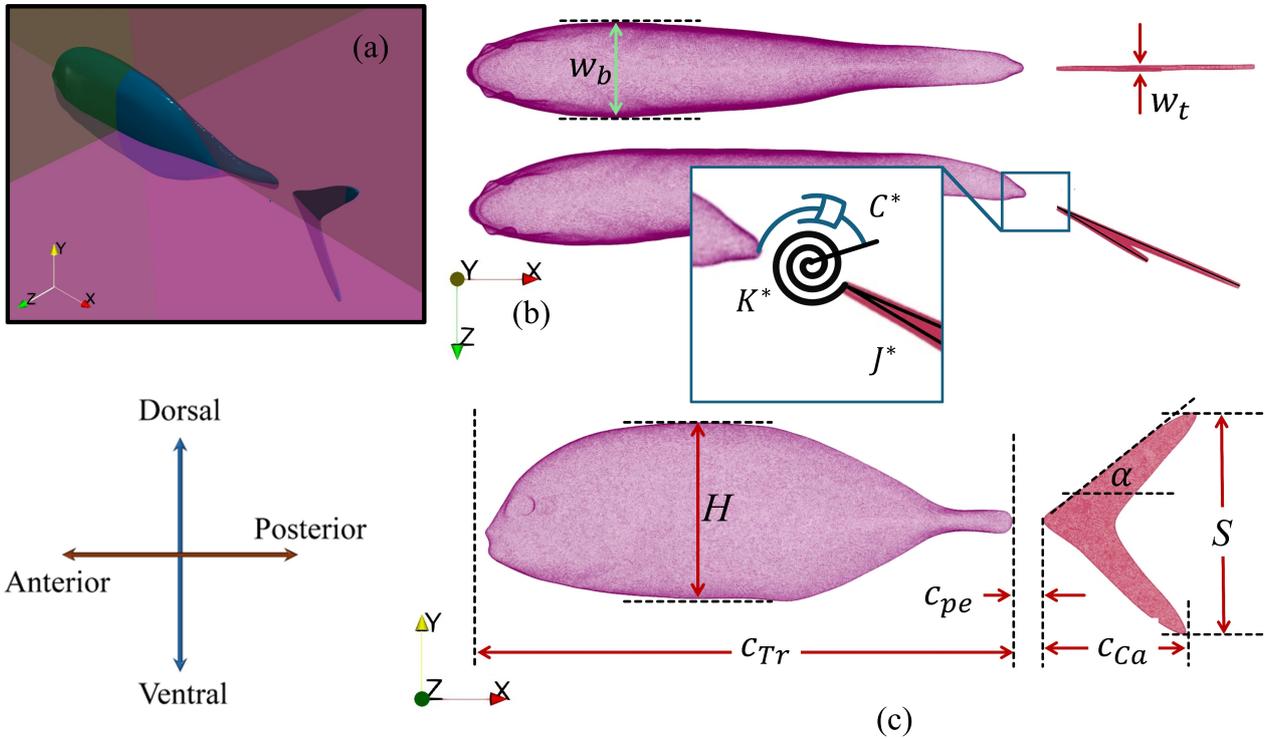}
    \caption{\edt{Physiological details of the Jack fish-like swimmer} at (a) a static position (${t}/{\tau} =0$), (b) mid-oscillation instant (${t}/{\tau} =0.55$) \edt{and characterization of the body-tail joint}, \edt{and (c) its side view with dimensional features} }
    \label{fig:geometry}
\end{figure}

We model the geometry of a Jackfish and its kinematics similar to that of reported \edt{by Khalid et al.} \cite{khalid2021larger}. \edt{However,} we modify the geometry of the \edt{swimmer} to construct the model with an independent caudal fin\edt{,} which is connected to the trunk of the Jackfish \edt{through a modeled joint}. Both the trunk and the caudal fin are scaled down by $13\%$ to ensure that the total length $L$ of the Jackfish remains unit, and the gap between the trunk and caudal fin is at $c_{pe}=0.05L$ as shown in Fig.~\ref{fig:geometry}. The gap between the trunk and the caudal fin alongside the length of the trunk ($c_{Tr}=0.75L$), and the caudal fin ($c_{Ca}=0.20L$) is chosen as reported by Gao et al. \cite{gao2018independent} and Nayal et al. \cite{nayak2025comparative}, \edt{while it serves the purpose of dealing with computational grid in our simulations}.

Figure.~\ref{fig:geometry} shows the flexure of the trunk with a passively pitching caudal fin \edt{that also performs prescribed heaving to conserve the natural undulatory kinematic profile}. The caudal fin is attached to the trunk using a torsional spring and a viscous damper at the peduncle\edt{,} as illustrated in Fig.~\ref{fig:geometry}b. \edt{Figures}~\ref{fig:geometry}b and \ref{fig:geometry}c \edt{show} important parametric quantities \edt{utilized in} this study. \edt{Please refer to T}able.~\ref{tab:geom_quantities} for the normalized \edt{morphological} parameters specified for the \edt{swimmer}. Here, the reference area of the caudal fin is calculated based on the \edt{aspect ratio ($AR$)} as a function of the span \edt{($S$)} of the trailing edge of the caudal fin ($AR = S^2/A_{CF}$)\edt{,} where $A_{CF}$ refers to the area of the caudal fin.

\begin{table}[ht!]
  \centering
   \caption{Geometric quantities of the Jackfish model with an independent caudal fin as a rigid body (all the quantities are normalized using using $L$)}
  \begin{tabular}{lccccccc}
    \toprule
    \textbf{$C_{Tr}$} & \textbf{$C_{pe}$} & \textbf{$C_{Ca}$} & \edt{\textbf{$\alpha~[^\circ]$}} & \textbf{$S$} & \textbf{$AR$} & \textbf{$W_b$} & \textbf{$w_t$}\\
    \midrule
    $0.75L$ & $0.05L$ & $0.20L$ & $40$ & $0.295L$ & $4.31$ & $0.12L$ & $0.005L$ \\
    \bottomrule
  \end{tabular}
   
  \label{tab:geom_quantities}
\end{table}

The trunk follows a carangiform undulating swimming flexure about \edt{z}-axis, with a continuous passive pitching \edt{and prescribed (active) heaving} of the caudal fin. The amplitude ($A\left(\frac{x}{L}\right)$) of the carangiform undulation is described by the following equation \cite{khalid2020flow, khalid2021larger, khalid2016hydrodynamics}.

\begin{align}
A\left(\frac{x}{L}\right) &= 0.02 - 0.085\left(\frac{x}{L}\right) + 0.1625\left(\frac{x}{L}\right)^2; \,\,\,\, 0 < \frac{x}{L} < c_b
\label{eq:eq1}
\end{align}

\edt{The swimmer performs undulation with a frequency $f$, and the tail follows a continuous trajectory of the body by heaving at the undulation frequency, and tracing the displacement ($h$) of the virtual joint at its peduncle, $h = A(0.78)$. The oscillating motion and the undulating motion is modeled by the mathematical expression, as presented as Eq.~\ref{eq:eqmathModel}. The motion of the body and tail is initiated using a hyperbolic tangent function \edt{($g(t)$),} also referred to as a Sigmoid function given in Eq.~\ref{eq:hyper}, to ensure a smooth transition to the peak amplitude of undulation \edt{and active heaving and pitching}. In this formulation, \edt{there is a cubic power to the exponent to ensure} a faster yet smoother rise to the peak amplitude. Notably, applying the Sigmoid function enables the passively pitching tail to reach steady-state oscillation much faster than without it, thereby reducing the overall computational cost.

\begin{align}
z\left(\frac{x}{L}\right) &= g\left(t\right)\,A\left(\frac{x}{L}\right) \cos\left[2\pi\left({\frac{x}{\lambda}}-{f}{t}\right)\right]
\label{eq:eqmathModel}
\end{align}

\begin{align}
g\left(t\right) &= \left[\frac{e^{3t} - 1}{e^{3t} + 1}\right]
\label{eq:hyper}
\end{align}
}
\edt{Getting inspiration from} the stiffness-modulation mechanisms observed in biological peduncles, the passive pitching of the caudal fin is modeled using a \edt{linear damper and a torsional spring, possessing both linear and cubically nonlinear characteristics}. As demonstrated by Lu et al.~\cite{lu2025nonlinear}, the musculature and tendon structures in the caudal peduncle do not behave as linear elastic elements\edt{. Instead}, they exhibit a displacement-dependent stiffness in which the restoring moment grows disproportionately at larger deflections. This nonlinear behavior allows real swimmers to sustain large pitching amplitudes while stabilizing the fin near its extreme \edt{angular positions} through a recoil-type response generated by the increased stiffness. Following this principle, we incorporate a cubic nonlinear stiffness term into the torsional spring model. The linear \edt{stiffness} term contributes to amplifying the pitching amplitude around the neutral position, while the cubic term acts as a stabilizing mechanism that limits excessive rotation by increasing the restoring torque at large angular excursions. This combination reproduces the characteristic ``oft-near-neutral, stiff-near-extremes'' behavior associated with nonlinear peduncle dynamics and supports synchronized flapping. The resulting passive pitching \edt{dynamics} of the caudal fin is governed by the following equation (Eq.~\ref{eq:non}).

\begin{equation}
    \left[\frac{2 \, c_{t}^{2}}{U_{\infty}^{2}} \, J^{\ast}\right] \ddot{\theta}_p+\left[\frac{2 \, c_{t}}{U_{\infty}} \, C^{\ast}\right] \dot{\theta}_p-\left[2 \, K^{\ast}\right] (A\theta_p-B\theta_p^3)=C_{M}
    \label{eq:non}
\end{equation}

\[
J^{\ast} = \frac{J_\theta}{\rho\,A_{ref}\,c_t^{3}},\;
K^{\ast} = \frac{K_\theta}{\rho\,U_{\infty}^{2}\,A_{ref}\,c_t},\;
C^{\ast} = \frac{c_\theta}{\rho\,U_{\infty}\,A_{ref}\,c_t^{2}},\;
C_{M} = \frac{2M}{\rho\,U_{\infty}^{2}\,A_{ref}\,c_t}.
\]

\noindent \edt{where} $\rho$ denotes \edt{density of} the fluid, \edt{and} $U_{\infty}$ is the freestream velocity. The term \edt{$C_M$} represents the hydrodynamic moment \edt{coefficient} acting about the pitching axis located at the peduncle, while $J_{\theta}$ is the mass moment of inertia about this axis. The coefficients $c_{\theta}$ and $K_{\theta}$ correspond to the torsional viscous damping and torsional stiffness of the \edt{body-tail joint}, respectively. The parameters $A$ and $B$ \edt{in the restoring term} describe the relative contributions of the linear and nonlinear components of the torsional stiffness. They satisfy the constraint $A + B = 1$, such that the restoring moment $K_{\theta}(A\theta_p - B\theta_p^3)$ represents a distributed combination of linear ($A\theta_p$) and \edt{cubically} nonlinear ($B\theta_p^3$) stiffness. This formulation allows the joint to behave as a soft spring near the neutral position while providing increasing resistance at larger angular amplitudes, consistent with nonlinear peduncle mechanics observed in biological swimmers. The dimensionless stiffness coefficient $K^{\ast}$ is treated as a dependent variable that varies with the dimensionless inertia $J^{\ast}$, as detailed in Eq.~\ref{eq:dimlessStiffnessStiff}.

\begin{equation}
    K^{\ast}= J^{\ast} \left[\frac{\pi\, n \, c_t \, f^{\ast}}{a}\right]^2 
    \label{eq:dimlessStiffnessStiff}
\end{equation}

\begin{equation}
    f^{\ast} = \frac{{2af} }{U_{\infty}}
    \label{eq:dimlessStiffnessFrequency}
\end{equation}

In Eqs.~\ref{eq:dimlessStiffnessStiff} and \ref{eq:dimlessStiffnessFrequency}, $a$ denotes the tail-beat amplitude of the body measured at $x/L = 1.0$ for this part of the analysis, and $f$ is the frequency of undulation of the swimmer. The parameter $n$ is referred as a synchronizing parameter to tune the pitching of the caudal fin with the undulation for incorporating the flapping of the caudal fin \cite{sridhar1975nonlinear, baek2001response}. For this study, \edt{a number of simulations are carried out} with the different values of $n$ alongside a range of values selected for the damping ratio ($\zeta$). The normalized damping coefficient $C^{\ast}$ is determined from the damping ratio $\zeta$ and the dimensionless stiffness $K^{\ast}$ according to Eq.~\ref{eq:damping}, following the standard formulation for nonlinear oscillators \cite{kim2018effect}.

\begin{equation}
    C^{\ast} = 2 \, \zeta\,\sqrt{J^{\ast} K^{\ast}}
    \label{eq:damping}
\end{equation}

\subsection{Governing equations for fluid flows}

For the current study, we use OpenFoam/v$2312$, which is an open source CFD solver \edt{and} offers \edt{a} variety of numerical techniques to compute different terms in the governing equations for the fluid flows. Hence, we directly solve the unsteady Navier-Stokes equations for the three-dimensional \edt{incompressible} fluid flows around the swimmer. To \edt{investigate} the independently pitching tail, we incorporate multiple variables for this study \edt{and} conduct $45$ \edt{three-dimensional high-fidelity} simulations. \edt{Our} analysis provides a strong foundation for \edt{investigating the influence of relevant dynamic parameters on the swimmer's} hydrodynamic performance. The non-dimensional forms of the continuity and incompressible Navier-Stokes equations are given below \cite{kamran2024does, khalid2016hydrodynamics}:

\begin{equation}
    \frac{\partial u_{j}}{\partial x_{j}}=0
\label{eq:eqCont}
\end{equation}

\begin{equation}
    \frac{\partial u_{i}}{\partial t}+\frac{\partial}{\partial x_{j}}\left(u_{i} u_{j}\right)=-\frac{1}{\rho} \frac{\partial p}{\partial x_{i}}+\nu \frac{\partial^{2} u_{i}}{\partial x_{j} \partial x_{j}}
\label{eq:eqNavStoke}
\end{equation}

\noindent \edt{where} $i,j = {1,2} $, $u_i$ and $u_j$ are the Cartesian components of the flow velocity, $p$ is the pressure, and $\rho$ is the density of the fluid. The temporal term in the governing equations are discretized using an implicit backward difference scheme. The $PIMPLE$ algorithm is used to couple the pressure and velocity field over the moving mesh. This algorithm combines the Pressure-Implicit with Splitting of Operators ($PISO$) algorithm and Semi-Implicit Method for Pressure-Linked Equations ($SIMPLE$) algorithm. The convergence criterion for the iterative solution at each time step is set to $10^{-04}$. In this study, a Laplace equation with inverse-distance diffusivity is used for dynamic meshing \cite{jasak2009dynamic}.

\begin{figure}[ht!]
    \centering
    \includegraphics[width=1\linewidth]{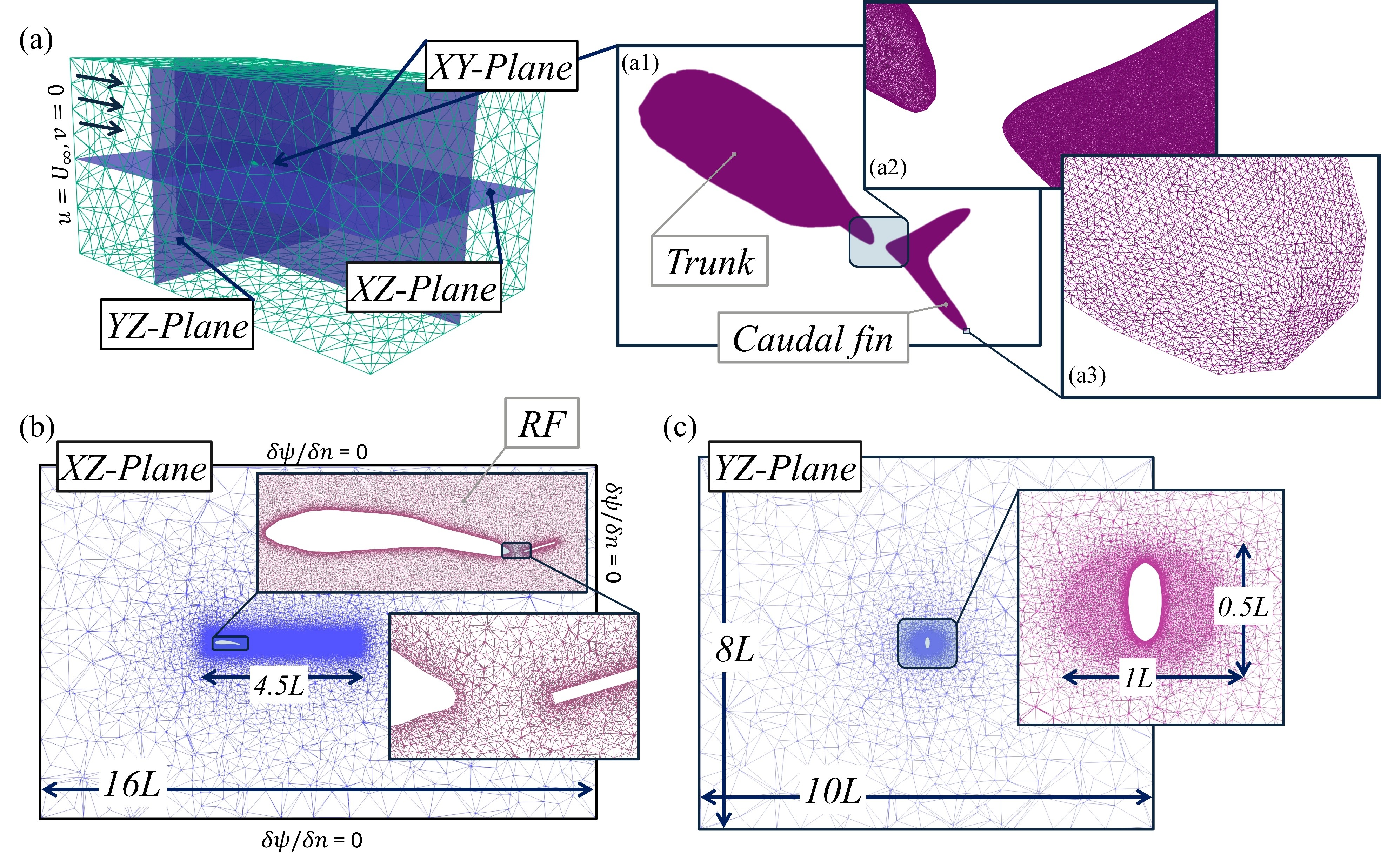}
    \caption{\dev{Computational domain and boundary conditions. (a) Three-dimensional domain with XY-, XZ-, and YZ-planes, (a1) fish geometry (trunk and caudal fin), and (a2)-(a3) surface mesh. (b) XZ-plane showing domain extent ($16L$) and downstream wake refinement ($RF$) ($4.5L$). (c) YZ-plane showing cross-sectional domain ($10L \times 8L$) and local mesh refinement around the swimmer.}}
    \label{fig:myMesh}
\end{figure}

An unstructured grid in a three-dimensional cuboid domain is used in \edt{our present simulations}. The grid illustrated in Fig.~\ref{fig:myMesh} have boundaries selected to ensure minimal numerical errors. \edt{To accurately handle the swimmer's motion,} the computational grid deforms at every time step by solving a displacement Laplacian equation, $\nabla \cdot (\gamma \nabla \xi) = 0$. A quadratic inverse-face-distance diffusivity ($\gamma=1/\sqrt{d}$) \edt{with reference to} the trunk and the caudal fin enforces near-rigid motion close to the bodies. \edt{This strategy significantly helps dissipation occur smoothly} away from them \cite{jasak2009dynamic}. The dimensions of the grid are selected to mitigate the cell displacement to ensure that the moving mesh does not influence the numerical accuracy.

\subsection{Verification \& validation}


\begin{figure}[ht!]
    \centering
    \includegraphics[width=1\linewidth]{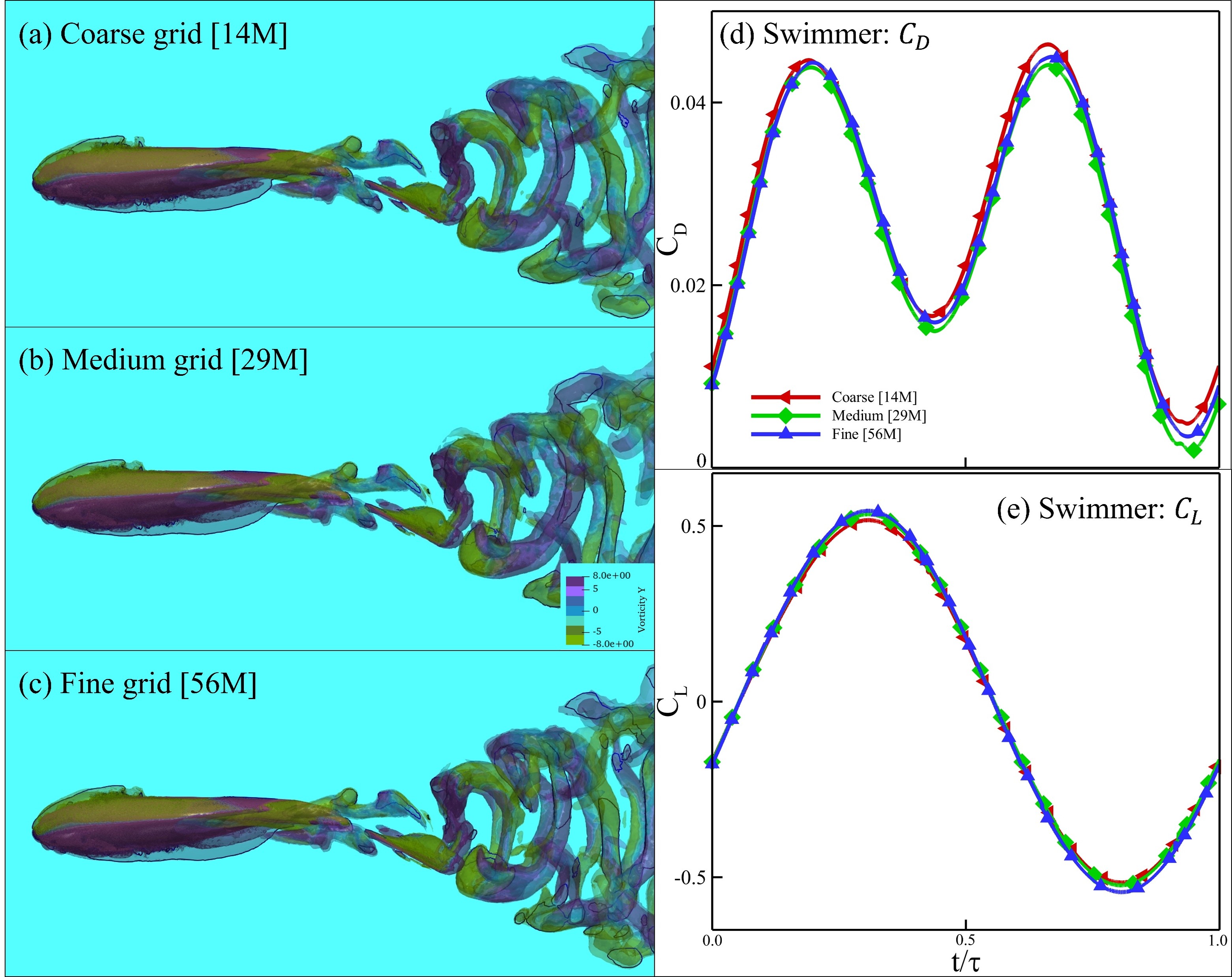}
    \caption{\dev{Grid-independence study. Instantaneous vorticity fields for (a) coarse ($14$M), (b) medium ($29$M), and (c) fine ($56$M) grids. Time histories of (d) drag coefficient $C_D$ and (e) lift coefficient $C_L$ of the swimmer over one oscillation cycle, respectively.}}
    \label{fig:gridvalidation}
\end{figure}

We \edt{choose} three grid resolutions to ensure the verification of our computational setup. For the grid-convergence study, \edt{we use three different grids of varying resolution at $\mathrm{Re}=3000$, and $f^{\ast}=0.60$ for flows over an actively pitching caudal fin. Following are the details for the three grids}: Grid $1$ with $14$M cells (\edt{c}oarse \edt{g}rid), Grid $2$ with $29$M cells (\edt{m}edium \edt{g}rid), and Grid $3$ with $56$M cells (\edt{f}ine \edt{g}rid). The levels of refinement and \edt{a} comparison between the cell size near the boundaries are \edt{presented} in Table~\ref{tab:mesh_refinement}. As \edt{shown} in \edt{Figs.~\ref{fig:gridvalidation}a, ~\ref{fig:gridvalidation}b, and \ref{fig:gridvalidation}c,} the dorsal view of the swimmer suggests no evident difference in the coherent structures in the wake of the swimmers \edt{obtained from} the coarse, medium, and fine grids. Furthermore, \edt{comparisons} between the drag coefficient ($C_D$) and the lift coefficient ($C_L$) from Figs.~\ref{fig:gridvalidation}d and \edt{\ref{fig:gridvalidation}}e, respectively, \edt{obtained from the three} grids \edt{indicate} that the coarse grid produces results that are nearly identical to those \edt{attained from} the medium and fine grids. Therefore, the coarse grid is selected for all subsequent simulations \edt{reported in this work}. For this study, we choose the value of the time step size ($\Delta t$) based on $7500$ time steps per oscillation cycle \edt{previously reported by} Nayak et al. \cite{nayak2025comparative}.

\begin{table}[ht!]
  \centering
  \caption{Details for mesh refinements in coarse, medium, and fine grids\edt{, where} Refinement (I) indicates \edt{a} comparison between \edt{the} coarse and medium grids, while Refinement (II) indicates \edt{a} comparison between \edt{the m}edium and \edt{f}ine grids. 
  \dev{The cell sizes around the trunk, caudal fin, and refinement region (RF) for the coarse, medium, and fine grids.}}
  \begin{tabular}{lcccccc}
    \toprule
    \textbf{-} & \textbf{Coarse} & \textbf{Medium} & \textbf{Refinement (I)} & \textbf{Medium} & \textbf{Fine} & \textbf{Refinement (II)} \\
    \midrule
    Cells & 13{,}764{,}344 & 28{,}665{,}094 & {52\%} & 28{,}665{,}094 & 55{,}629{,}584 & {48\%} \\
    \dev{Trunk} & 0.0015 & 0.0008 & {47\%} & 0.0008 & 0.00055 & {31\%} \\
    \dev{Caudal fin} & 0.0010 & 0.0006 & {40\%} & 0.0006 & 0.00035 & {42\%} \\
    RF   & 0.0085 & 0.0070 & {18\%} & 0.0070 & 0.0065 & {7\%} \\
    \bottomrule
  \end{tabular}
    
\label{tab:mesh_refinement}
\end{table}

\begin{figure}[ht!]
    \centering
    \includegraphics[width=1\linewidth]{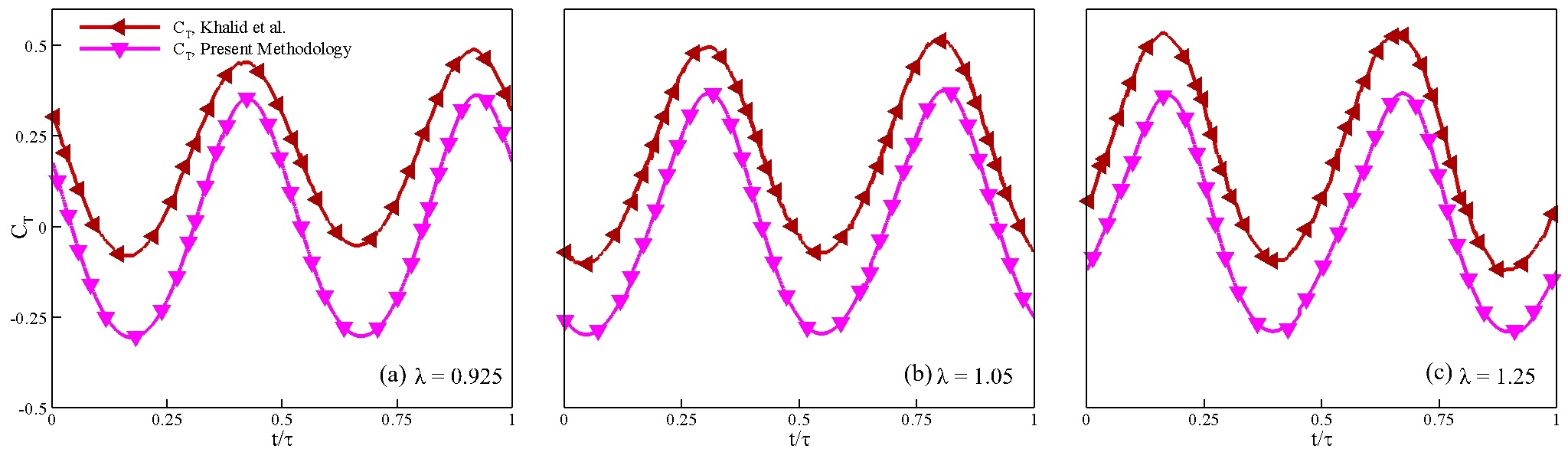}
    \caption{\dev{Validation of the thrust coefficient $C_T$ obtained using the present methodology against Khalid et al. \cite{khalid2021larger} for different wavelengths, (a) $\lambda = 0.925L$, (b) $\lambda = 1.05L$, and (c) $\lambda = 1.25L$ over one oscillation cycle.}}
    \label{fig:solverValidation3D}
\end{figure}

Our \edt{present} computational methodology and the strategy for the morphing mesh is validated by comparing the thrust coefficient ($C_T=-C_D$) \edt{from our simulations with those provided by Khalid et al. \cite{khalid2021larger}}. \edt{Three distinct cases are considered} here corresponding to three values of \edt{undulation} wavelengths \edt{$\lambda/L = 0.925$}, $1.05$, and $1.25$, as shown in Figs.~\ref{fig:solverValidation3D}a, \edt{\ref{fig:solverValidation3D}b, and \ref{fig:solverValidation3D}c}, respectively. The plots of $C_T$ show similar \edt{trends} to \edt{the} reference case, \edt{and} the peaks in the oscillation cycle occur in the first half of the oscillation cycle for $\lambda = 0.925L$ as shown in Fig.~\ref{fig:solverValidation3D}a. \edt{For the} larger wavelength\edt{,} we observe that \edt{the peaks} in $C_T$ occurs in the second half of the oscillation cycle\edt{,} as shown in Figs.~\ref{fig:solverValidation3D}b and \edt{\ref{fig:solverValidation3D}c. The differences between the $C_T$ profiles arise due to the gap between the body and the tail in our simulations, which could slightly delay the interactions between the vortices shed from the body and the caudal fin.}

\newpage
\section{Results \& Discussion}

The \edt{chosen} kinematic parameters are \edt{presented in Table~\ref{tab:discussion_table_3D}} to ensure an effective design of simulations. The wavelength stays unchanged with $\lambda= 1.0L$\edt{, whereas} $f^\ast$, $\mbox{Re}$, $J^\ast$, $A$, and $B$ are fixed with their corresponding values shown in Table~\ref{tab:discussion_table_3D}. \edt{Moreover,} $\zeta$, and $n$ are varied with increments of $0.0167$, and $2$, respectively.

\begin{table}[ht!]
  \centering
  \caption{Governing parameters for the $3D$ analysis.}
  \label{tab:discussion_table_3D}
  \begin{tabular}{llc}
    \toprule
    Parameter & Symbol & Value \\
    \midrule
    Undulatory gait              & —               & Carangiform \\
    Strouhal frequency           & $f^{\ast}$            & $0.3$ \\
    Wavelength           & $\lambda$            & $1.0L$ \\
    Damping ratio                    & $\zeta$   & $0.300$-$0.450$ \\
    Dimensionless inertia & $J^{\ast}$   & $3.715$ \\
    Reynolds number              & Re            & $3000$ \\
    Tuning parameter              & $n$            & $42$-$60$ \\
    Coefficient of linear stiffness              & $A$            & $0.85$ \\
    Coefficient of non-linear stiffness              & $B$            & $0.15$ \\

    \bottomrule
  \end{tabular}
\end{table}

\begin{figure}[ht!]
    \centering
    \includegraphics[width=1\linewidth]{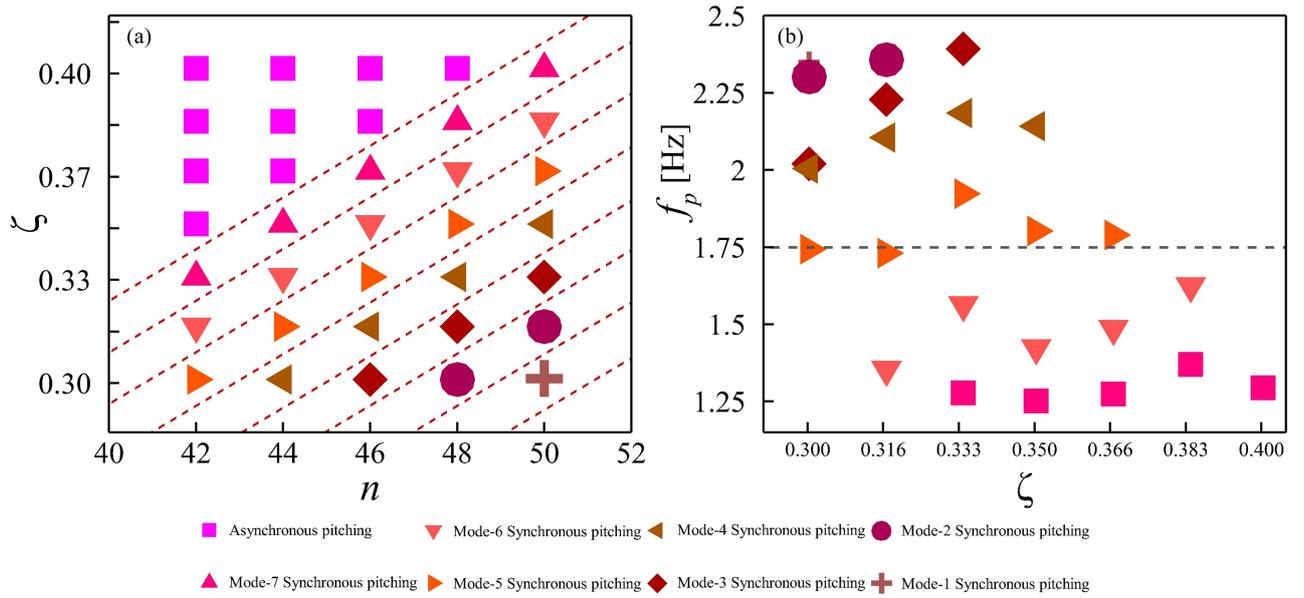}
    \caption{\dev{Synchronization map of passively pitching tail, (a) variation of the synchronization parameter $\zeta$ with tuning parameter $n$, showing asynchronous and synchronous pitching regimes (Modes~1--7). Dashed lines indicate constant phase-locking trends. (b) Corresponding pitching frequency $f_p$ as a function of $\zeta$, with the dashed line denoting the active heaving frequency.}}
    \label{fig:syncMap}
\end{figure}

\edt{In this study}, we observe that the linear stiffness acts as a negative recoil term, which is responsible for incorporating pitching with larger amplitude. \edt{However}, when the pitching approaches a large angle, the cubic non-linear stiffness becomes dominant and triggers the recoil, helping the fin return close to the neutral axis. From the test runs, varying $n$ led to tuning the pitching frequency close to the heaving frequency in correlation to the damping ratio. \edt{Importantly,} $n$ and $\zeta$ yield $35$ simulations, \edt{and the dynamical responses of the caudal fin are categorized on $\zeta-n$ and $f_p-\zeta$ planes} in Fig.~\ref{fig:syncMap}.

\edt{For reference, the undulation frequency of the trunk and the heaving frequency of the caudal fin are $1.75~\mbox{Hz}$} ($f^*=0.35$). Figure~\ref{fig:syncMap}a provides an overview of how $n$ and $\zeta$ correspond to the dynamic behavior of the caudal fin. The blue squares indicate pitching \edt{asynchronous to heaving of the caudal fin}, characterized by irregular or unilateral pitching. These asynchronous \edt{motion-related} cases are excluded from \edt{our} further analysis in this study. Modes~1–7 represent categories describing how pitching of the caudal fin evolves with increasing $n$ and $\zeta$ linearly. Although simulations within each mode show variations in the fundamental pitching frequency ($f_p$), the discrepancies within a given trend remain relatively small, as illustrated in Fig.~\ref{fig:syncMap}b. This figure shows that, among all simulations performed, fundamental frequency of the simulations from  Mode~5 aligns most closely with the undulation frequency, which is highlighted using a dashed gray line across the plot. Moving away from Mode~5, either by increasing or decreasing $\zeta$, the fundamental frequency of pitching deviates significantly from the undulation frequency. To investigate this behavior further, we extend the analysis by running an additional set of five simulations following the linear trend of Mode~5, resulting in a total of ten simulations considered for subsequent analysis, the configurations of the simulations are listed in Table~\ref{tab:mode_5}.

\begin{table}[ht!]
  \centering
  \caption{{Parameters for the extended analysis of} Mode-5 synchronous \edt{cases} for the pitching of the caudal fin.}
  \begin{tabular}{lcccccccccc}
    \toprule
    \textbf{Cases} & {1} & {2} & {3} & {4} & {5} & {6} & {7} & {8} & {9} & {10}  \\
    \midrule
    \textbf{$\zeta$} & {0.300} & {0.317} & {0.333} & {0.350} & {0.367} & {0.383} & {0.400} & {0.417} & {0.433} & {0.450}  \\
    \textbf{$n$} & {42} & {44} & {46} & {48} & {50} & {52} & {54} & {56} & {58} & {60}  \\
    \bottomrule
  \end{tabular}
  
  \label{tab:mode_5}
\end{table}

From the cases listed in Table~\ref{tab:mode_5}, the steady state response is declared based on the stabilization of the pitching frequency of the caudal fin, which also corresponds to the fundamental \edt{natural} frequency \edt{in this motion}. The frequency of each pitching oscillation is recorded, and the deviation ($\Delta$) is calculated based on the frequency of the preceding oscillation, \edt{according to} the \edt{relation provided} below. 

\begin{equation}
\text{$\Delta$} = \left|\frac{f_{p_i}-f_{p_{i-1}}}{f_{p_i}}\right| \times 100\%
    \label{eq:deviation}
\end{equation}

\begin{figure}[ht!]
    \centering
    \includegraphics[width=1\linewidth]{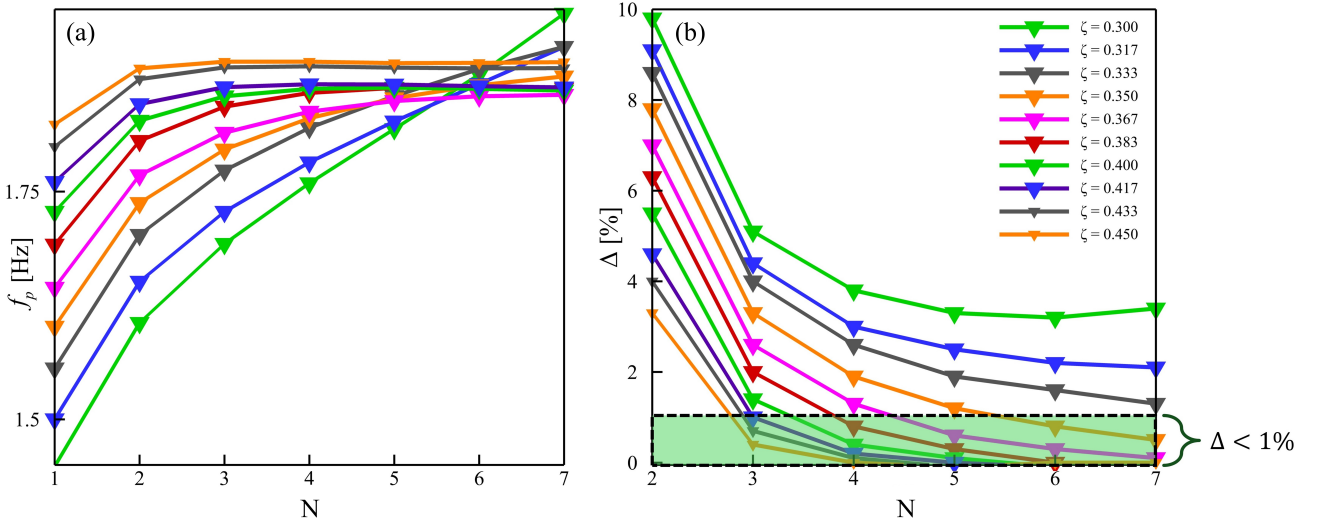}
    \caption{The plot shows the (a) cycle-by-cycle pitching frequency of the caudal fin for $\zeta=0.300$--$0.450$ with its corresponding tuning parameter $n$, \edt{and} (b) deviation of the frequency \edt{from the} preceding cycle till it reaches \edt{steady state}. }
    \label{fig:mode5}
\end{figure}

\noindent
Here, $f_{p_i}$ denotes the pitching frequency of the current cycle, and $f_{p_{i-1}}$ is the frequency from the preceding cycle. When the deviation falls within $1\%$, the cycle is considered stable, indicating that the pitching of the caudal fin \edt{settles} and \edt{reaches} its \edt{steady value}. The cycle-by-cycle frequencies for each case, along with their corresponding $\Delta$, are presented in Figs.~\ref{fig:mode5}a and \edt{\ref{fig:mode5}}b, respectively.

In Fig.~\ref{fig:mode5}a, a clear trend emerges between $f_p$ and $\zeta$. The pitching frequency reaches its steady state value within approximately three oscillation cycles. \edt{The} cases with \edt{a} lower $\zeta$ begin with a significantly reduced initial pitching frequency and gradually increases over successive cycles. \edt{Contrarily}, \edt{the} cases with higher $\zeta$ start at frequencies already close to their steady state values. This behavior is further illustrated in Fig.~\ref{fig:mode5}b, where cases with a low $\zeta$ exhibit an initial deviation of nearly $10\%$ from their actual pitching frequency, whereas cases with higher $\zeta$ deviate by less than $4\%$, followed by a rapid drop to below $1\%$. These observations suggest that, when synchronizing the caudal fin with the heaving through a nonlinear stiffness, the system reaches steady state more \edt{rapidly} with an appropriate choice of $\zeta$, allowing accurate prediction of \edt{the fin's} behavior from the first oscillation.

\begin{figure}[ht!]
    \centering
    \includegraphics[width=1\linewidth]{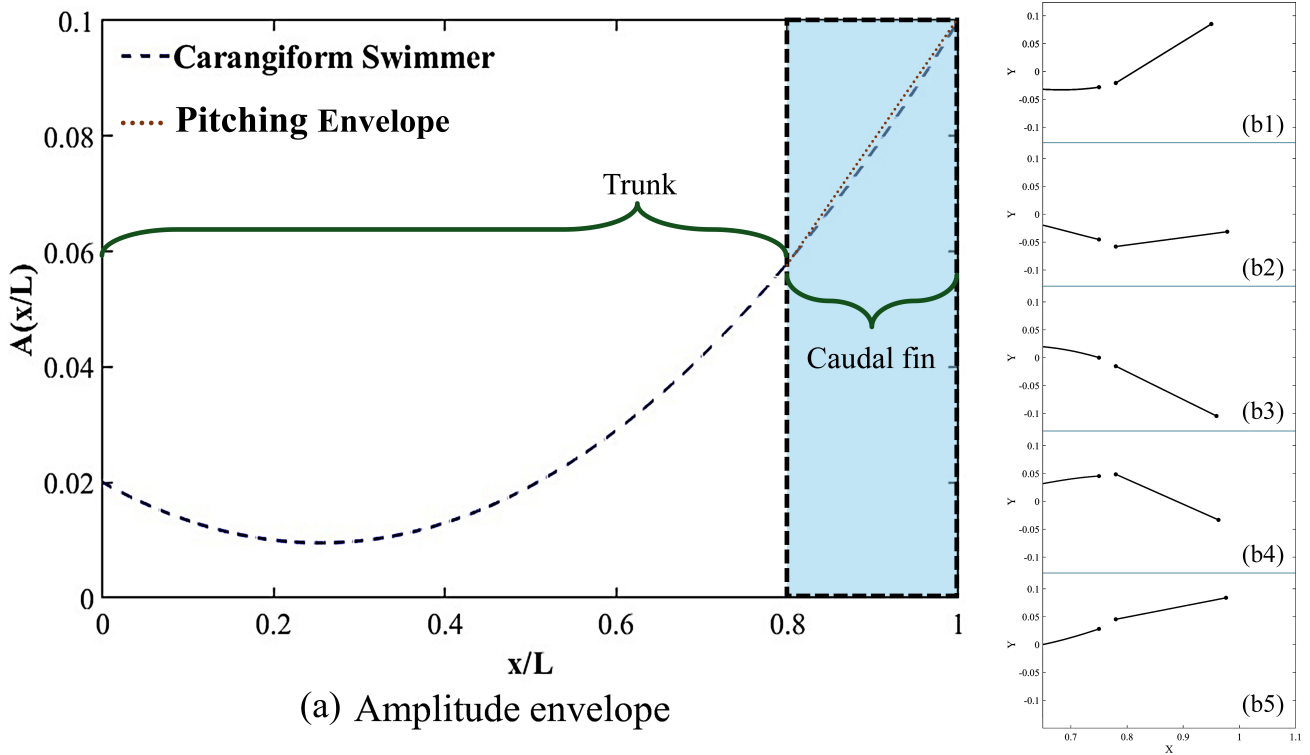}
    \caption{Amplitude envelope of \edt{the} carangiform swimmer with (a) undulatory \edt{profile} of the caudal fin from \edt{carangiform style} \edt{and} its pitching trajectory\edt{, and} (b) the active pitching trajectory of the caudal fin over a full oscillation cycle.}
    \label{fig:activePitching}
\end{figure}

To evaluate the performance of the configuration with a passively pitching caudal fin, we construct a reference model with an actively pitching tail. \edt{The purpose is to ensure} that the trailing edge of the caudal fin attains the maximum amplitude, $A(1.0)=0.1L$, matching the amplitude envelope of a carangiform swimmer, as shown in Fig.~\ref{fig:activePitching}a. \edt{This} figure illustrates the continuous amplitude distribution along the body of the swimmer, where the caudal fin occupies the region from $x/L=0.8$ to $1.0$ and spans a length of $0.2L$. \edt{Figures}~\ref{fig:activePitching}b$1$ and~\ref{fig:activePitching}b$5$ \edt{show} \edt{instantaneous} snapshots over a full oscillation cycle for the actively pitching caudal fin. The prescribed pitching is defined by the equation below:

\begin{equation}
\theta_{p}(t)=f\left(t\right)\,\sin^{-1}\!\Biggl\{
    \frac{1}{c_{t}}\Bigl[
        A(1.0)\,\cos\!\Bigl(2\pi\bigl(\tfrac{1.0}{\lambda}-ft\bigr)\Bigr) - A(0.8)\,\cos\!\Bigl(2\pi\bigl(\tfrac{0.8}{\lambda}-ft\bigr)\Bigr)
    \Bigr]
\Biggr\}.
\label{eq:active}
\end{equation}

\begin{figure}[ht!]
    \centering
    \includegraphics[width=1\linewidth]{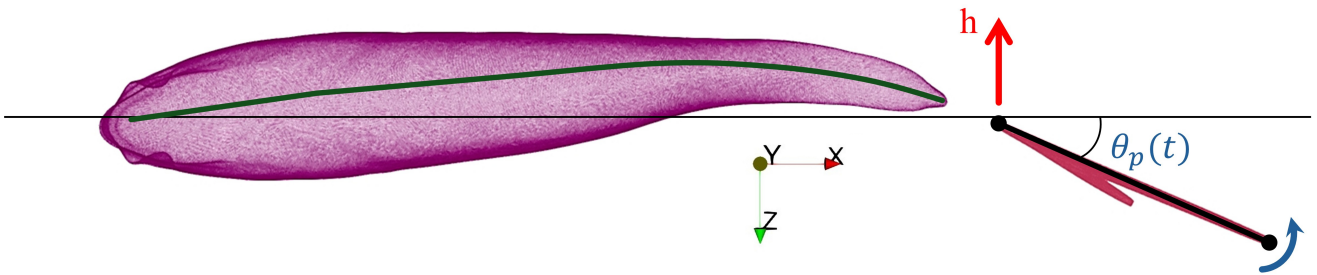}
    \caption{Schematic of Jackfish used to represent the pitching of the caudal fin in this study.}
    \label{fig:activeJackfish}
\end{figure}

\noindent \edt{where} $\theta_{p}(t)$ represents the pitching angle of the caudal fin, formulated to replicate the characteristic undulation of a carangiform swimmer. The \edt{schematics} in \edt{Figs}.~\ref{fig:activePitching}b \edt{are} derived from the Jackfish model shown in Fig.~\ref{fig:activeJackfish} and is used to visualize and analyze the trajectory traced by the caudal fin during swimming. In Fig.~\ref{fig:activeJackfish}, $h$ denotes the heaving displacement of the caudal fin, while $\theta_{p}(t)$ indicates its angular displacement.

\begin{figure}[ht!]
    \centering
    \includegraphics[width=1\linewidth]{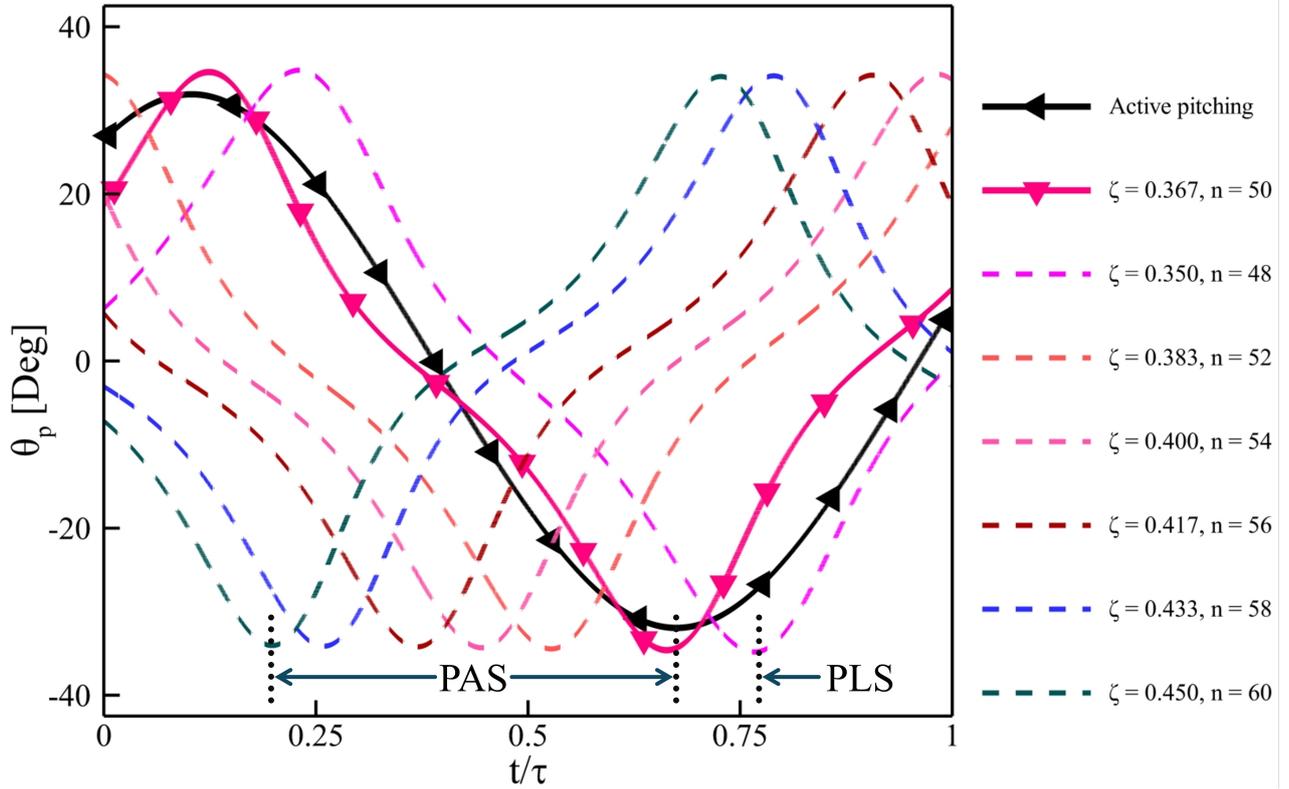}
    \caption{Pitching displacement of the caudal fin over a full oscillation cycle for the passively pitching tail alongside its active counterpart.}
    \label{fig:phaseBetweenOscillations}
\end{figure}

From the configurations presented in Fig.~\ref{fig:mode5}, the cases in which the passively pitching caudal fin reaches stability and steady states within seven oscillation cycles are examined further. The corresponding angular displacements \edt{for these cases} are plotted in Fig.~\ref{fig:phaseBetweenOscillations}. \edt{Here}, $\theta_{p}(t)$ for all stable passive pitching are displayed alongside a reference line in black line, that represents \edt{the case} with an actively pitching caudal fin. \edt{All} the configurations with a passively pitching caudal fin \edt{in Fig.~\ref{fig:phaseBetweenOscillations}} show a distinct behavior than that of the reference case \edt{with active pitching,} which shows the traditional Sinusoidal wave form with the maximum pitching angle of $32^\circ$. The cases with the passively pitching caudal fin demonstrates a sharp transition in the wave around the peaks \edt{that} correlates to the nonlinear stiffness stabilizing the system when the pitching approaches a large amplitude\edt{. It is important to notice that all the cases with passive pitching of the tail} exhibit \edt{larger angular} amplitudes than $32^\circ$. The configuration with $\zeta = 0.367$, and $n=50$ \edt{closely follow} the actively pitching reference case. \edt{For} the configuration with smaller and larger $\zeta$, the wave tends to display a lag, and advancement (lead), respectively. The phase lag synchronization ($\mbox{PLS}$) is referred to the wave which lags in comparison with the actively pitching case. The phase advanced synchronization ($\mbox{PAS}$) is referred to the wave \edt{that} leads in Fig.~\ref{fig:phaseBetweenOscillations}. With increasing $\zeta$ and $n$\edt{,} the wave exhibits higher $\mbox{PAS}$.

\begin{figure}[ht!]
    \centering
    \includegraphics[width=1\linewidth]{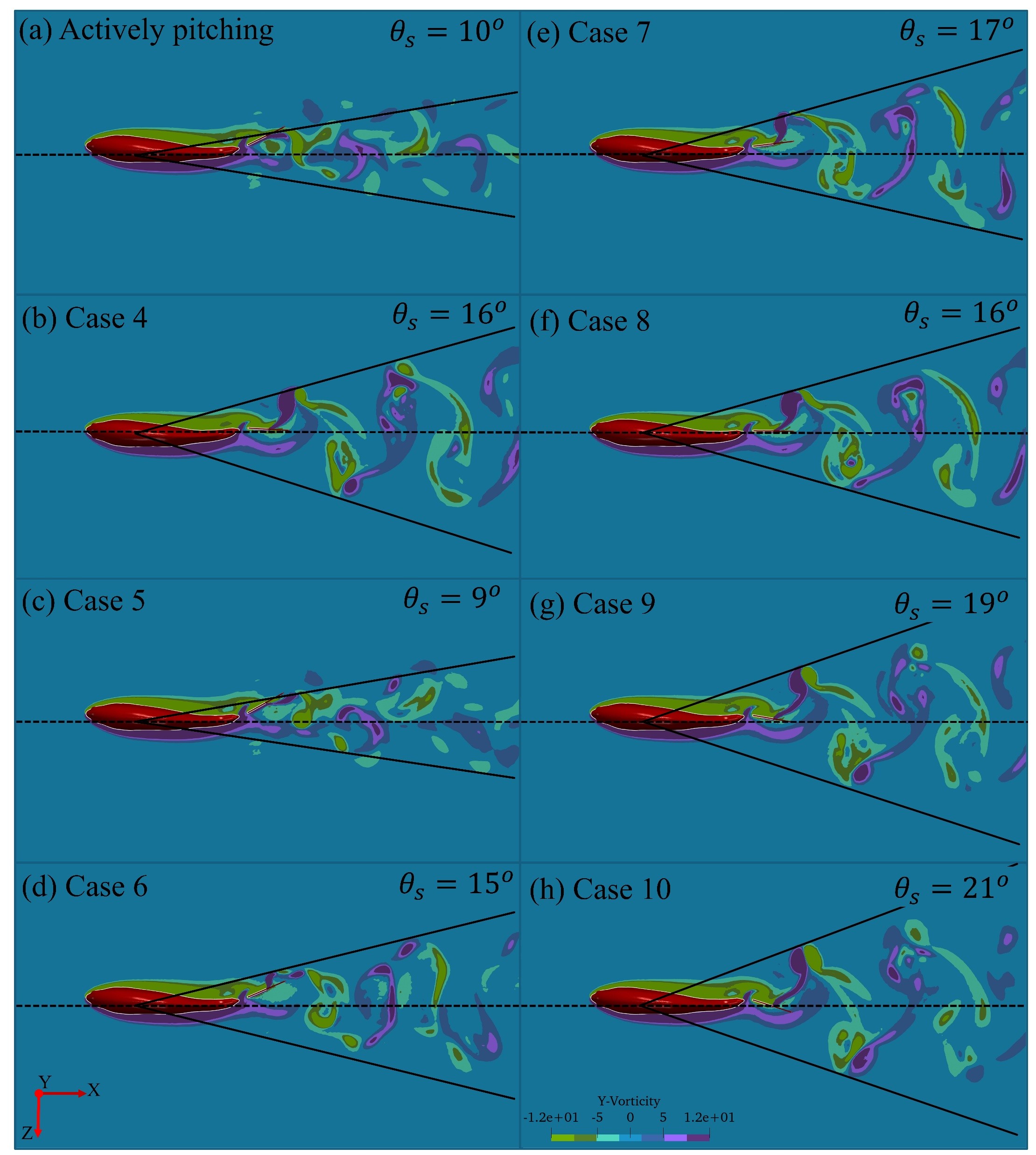}
    \caption{Wake signature for all of the configurations from a dorsal view. for (a) an actively pitching tail, alongside (b)-(h) the passively pitching tail for case $4$ to $10$, respectively.}
    \label{fig:vortexWakeAllCases}
\end{figure}

 \edt{Now, it becomes important and meaningful to} analyze the influence \edt{of} $\mbox{PAS}$ and $\mbox{PLS}$ on vortices generated in the wake of the swimmer. We plot \edt{contours of the out-of-plane vorticity component ($\omega_y$) by setting up} the swimmer's dorsal view for these \edt{stable} configurations in Fig.~\ref{fig:phaseBetweenOscillations}. \edt{Here, the wake patterns from the case of an} actively pitching caudal fin \edt{is also included}. For each case, we report the wake-spread angle $\theta_s$, which quantifies the deviation of the shed vortices from the streamwise direction relative to the centerline of the swimmer. The actively pitching configuration \edt{in Fig.~\ref{fig:vortexWakeAllCases}a} generates a narrow wake with $\theta_s \approx 10^\circ$ , indicating that the vortices are shed predominantly in the streamwise direction \edt{without experiencing large deflections}. For the passively \edt{pitching tails}, distinct trends emerge. In case~4 (\edt{see} Fig.~\ref{fig:vortexWakeAllCases}b), which displays a pronounced $\mbox{PLS}$ response, the wake-spread angle increases to approximately $16^\circ$, reflecting both a greater lateral displacement of the vortices and a broader spanwise spread. Case~5, whose kinematics most closely resembles the actively pitching fin, produces a comparatively smaller deflection. As we progress from case~6 through case~10 (Figs.~\ref{fig:vortexWakeAllCases}d–h), $\theta_s$ increases, correlating with the progressive intensification of $\mbox{PAS}$ \edt{for} these configurations.

Now, we further \edt{examine} the performance and the effect of the nonlinear stiffness \edt{of the body-caudal fin joint of the model} carangiform swimmer. For this purpose, \edt{we} select the cases (\edt{case 5}) {demonstrating the} behavior \edt{similar to} that of an actively pitching tail and \edt{another} case (\edt{case 9}) of \edt{the synchronously pitching and heaving fin} \edt{that} largely offsets from the reference case. \edt{Please note that case 5 and case 9 correspond} to \edt{minimum} and maximum offset from the actively pitching caudal fin, respectively. 

\begin{figure}[ht!]
    \centering
    \includegraphics[width=1\linewidth]{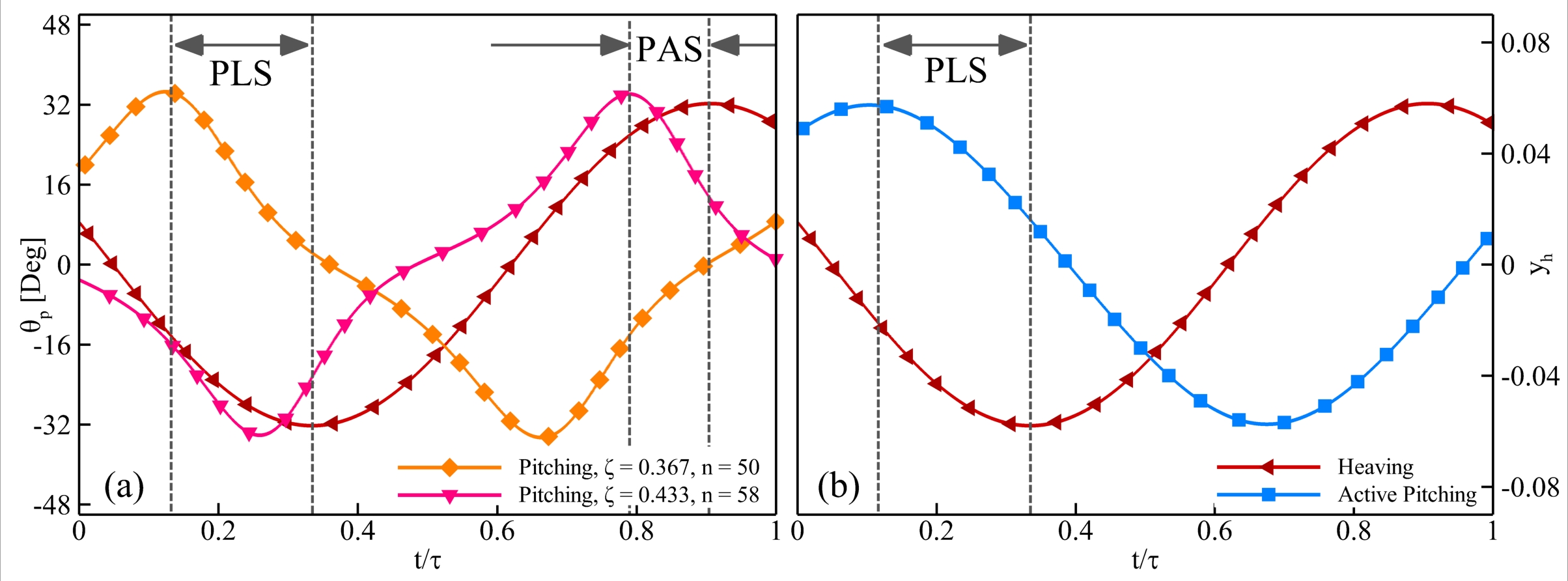}
    \caption{The comparison of the heaving and pitching of the caudal fin for (a) passively pitching tail with $\zeta = 0.367$ and $n = 50$, and $\zeta = 0.433$ and $n = 58$ corresponding to case $5$, and case $9$, respectively, \edt{and} (b) the actively pitching tail}
    \label{fig:waveComparison}
\end{figure}

Figure~\ref{fig:waveComparison} presents the angular displacement of the caudal fin together with its heaving displacement over one oscillation cycle. The heaving displacement is prescribed such that the leading edge of the caudal fin follows the trajectory of a carangiform swimmer, as discussed earlier. Figure~\ref{fig:waveComparison}a illustrates the pitching for cases $5$ and $9$ alongside the heaving, while Fig.~\ref{fig:waveComparison}b shows the corresponding displacement for the actively pitching tail alongside its heaving . In case $5$, the pitching displacement lags behind the heaving, which closely resembles the phase lag observed in the actively pitching case shown in Fig.~\ref{fig:waveComparison}b, indicating the presence of a $\mbox{PLS}$. Conversely, the pitching in case $9$ exhibits $\mbox{PAS}$, which is notably distinct from the actively pitching configuration.

\begin{figure}[ht!]
    \centering
    \includegraphics[width=1\linewidth]{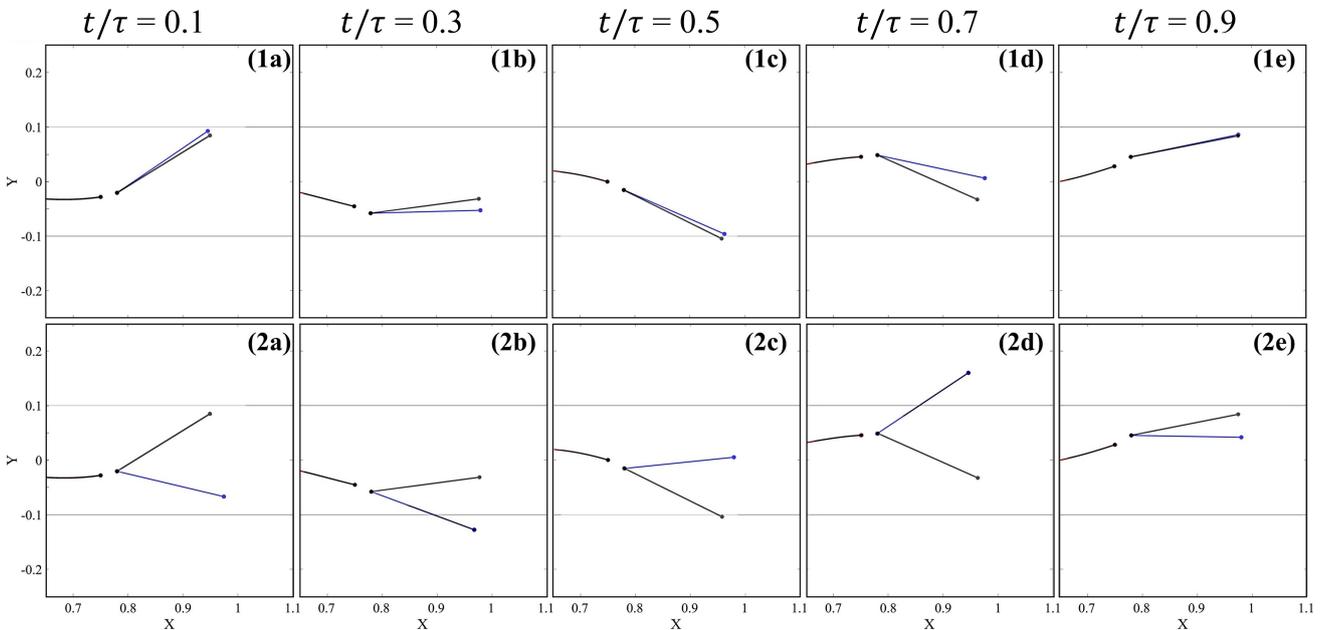}
    \caption{Schematic of the ($1a$)-($1e$) case $5$, and ($2a$)-($2e$) case $9$ at different stages of its full oscillation cycle. In comparison with the schematic of the actively pitching case in black.}
    \label{fig:case5_9}
\end{figure}

In Fig.~\ref{fig:case5_9}, the \edt{instantaneous positions of the caudal fin in} cases $5$ and $9$ are illustrated at five distinct \edt{time instances during a} complete oscillation cycle. \edt{Here, }the caudal fin undergoing passive pitching is shown by the blue line, whereas the actively pitching trajectory is shown in black. Case $5$, in Fig.~\ref{fig:case5_9}($1$a)–\ref{fig:case5_9}($1$e), exhibits a close agreement between the passive and active pitching throughout the cycle. The primary deviation occurs near the neutral axis, consistent with the phase disparity observed earlier in Fig.~\ref{fig:phaseBetweenOscillations}. Furthermore, for both passive and active configurations, the \edt{trailing edge ($\mbox{TE}$) of the caudal fin} traverses the same peak-to-peak amplitude envelope \edt{as that of a full undulation profile for the} carangiform swimmer. In contrast, case $9$ demonstrates a pronounced deviation between the two \edt{kinematics}. The passively pitching caudal fin reaches and departs from its peak positions abruptly, advancing ahead of the heaving. This sharp recoil decelerates as the fin approaches the neutral axis, resulting in a reduced phase difference between the passive and active pitching, as illustrated in Figs.~\ref{fig:case5_9}($2$b), \ref{fig:case5_9}($2$c), and \ref{fig:case5_9}($2$e). The pitching motion in cases $5$ and $9$ exhibit a comparable recoil behavior near the peak displacement, which exerts a dominant influence on both the \edt{streamwise force} coefficient ($C_D$) and the moment coefficient ($C_M$), as observed \edt{and explained} in the subsequent analysis.

\begin{figure}[ht!]
    \centering
    \includegraphics[width=1\linewidth]{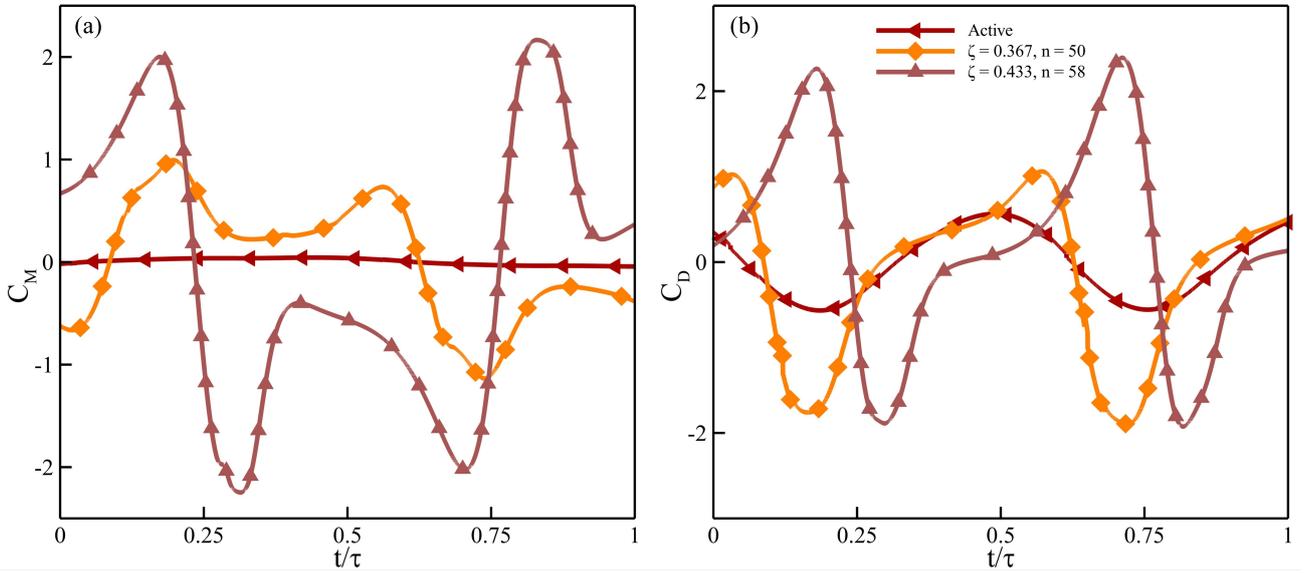}
    \caption{(a) \edt{Comparative plots for the} moment coefficient ($C_M$), and (b) drag coefficient ($C_D$) \edt{for} case $5$, case $9$, and the actively pitching tail over an oscillation cycle.}
    \label{fig:CDCM}
\end{figure}

The analysis of the moment and drag coefficients ($C_M$ and $C_D$) presented in Figs.~\ref{fig:CDCM}a and~b, respectively, provides deeper insight into the \edt{body-tail-fluid interactions}. As shown in Fig.~\ref{fig:CDCM}a, \edt{the amplitude of $C_M$} for case~$9$ is markedly higher than that of case~$5$ and the actively pitched configuration. Although case~$5$ exhibits a smaller $C_M$ compared to case~$9$, it remains higher than the actively pitching tail. Both passive and active pitching displays pronounced spikes in $C_M$ around the instant when the caudal fin abruptly recoils from the peak position. In Fig.~\ref{fig:CDCM}b, the variation of $C_D$ reveals that case~$5$ exhibits a thrust-dominant response, whereas the actively pitched caudal fin maintains a nearly symmetric profile. Conversely, case~$9$ exhibits large amplitude fluctuations in $C_D$, with the waveform remaining predominantly within the \edt{range of} positive $C_D$, \edt{indicating} a drag-dominated behavior. These observations suggest that the passively pitching configuration in case~$5$ offers a hydrodynamic advantage over both the highly asynchronous case~$9$ and the actively pitching counterpart. The contrast in $C_M$ and $C_D$ suggests distinct shedding of leading-edge vortices ($\mbox{LEV}$) and trailing-edge vortices $\mbox{TV}$ and \edt{their interactions for the scenarios considered here}. It prompts an analysis of the near-wake \edt{flow} structure and \edt{their} phase relation to the motion of the caudal fin.

\begin{figure}[ht!]
    \centering
    \includegraphics[width=1\linewidth]{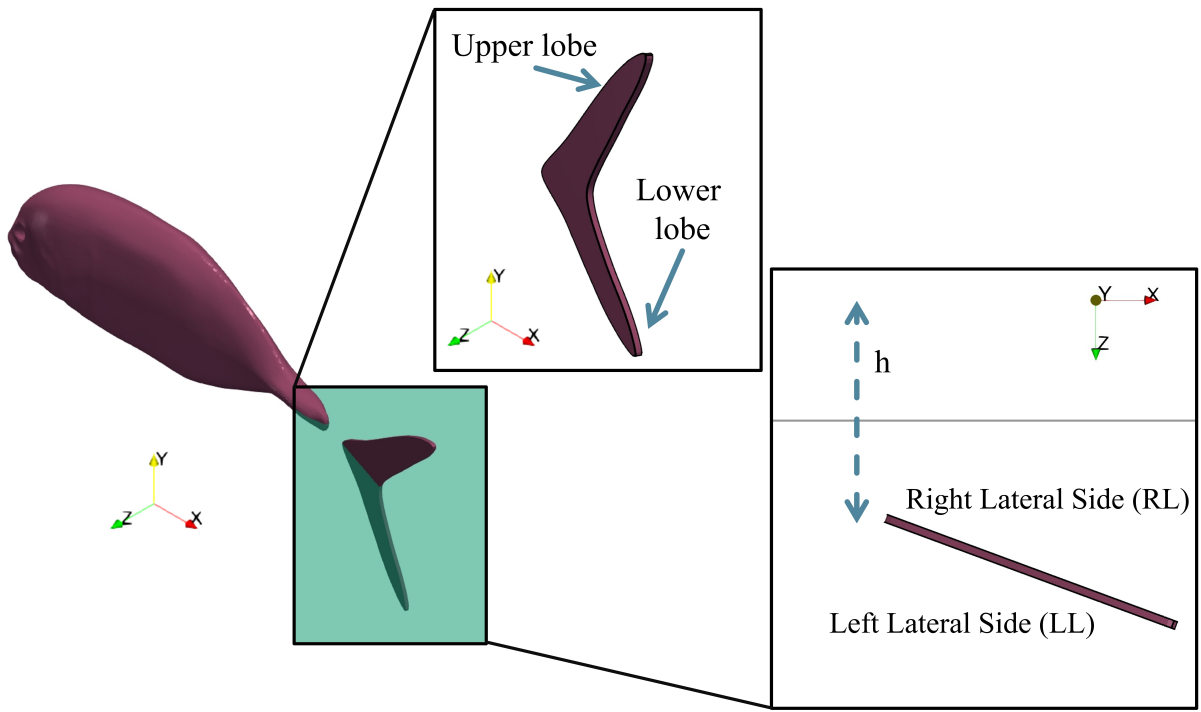}
    \caption{Nomenclature used to track the vortices developed and sheded from the caudal fin of the Jackfish.}
    \label{fig:NomenclaturePlot}
\end{figure}

To facilitate a clear and consistent interpretation of the hydrodynamic mechanisms \edt{governing the production of these hydrodynamics forces and moments}, a schematic representation of the caudal fin and associated terminology is provided in Fig.~\ref{fig:NomenclaturePlot}. This illustration describes the upper and lower lobes of the bi-lobed caudal fin, the right ventral and left ventral sides, and the heaving displacement. Establishing this nomenclature is essential for maintaining clarity in subsequent discussions, when particularly describing vortex formation, detachment, and their spatial relationship to the surface of the fin. By standardizing the reference frame and labeling conventions, the schematic ensures that the interpretation of the wake structures and their phase alignment with the fin's kinematics is both precise and unambiguous.

\begin{figure}[ht!]
    \centering
    \includegraphics[width=1\linewidth]{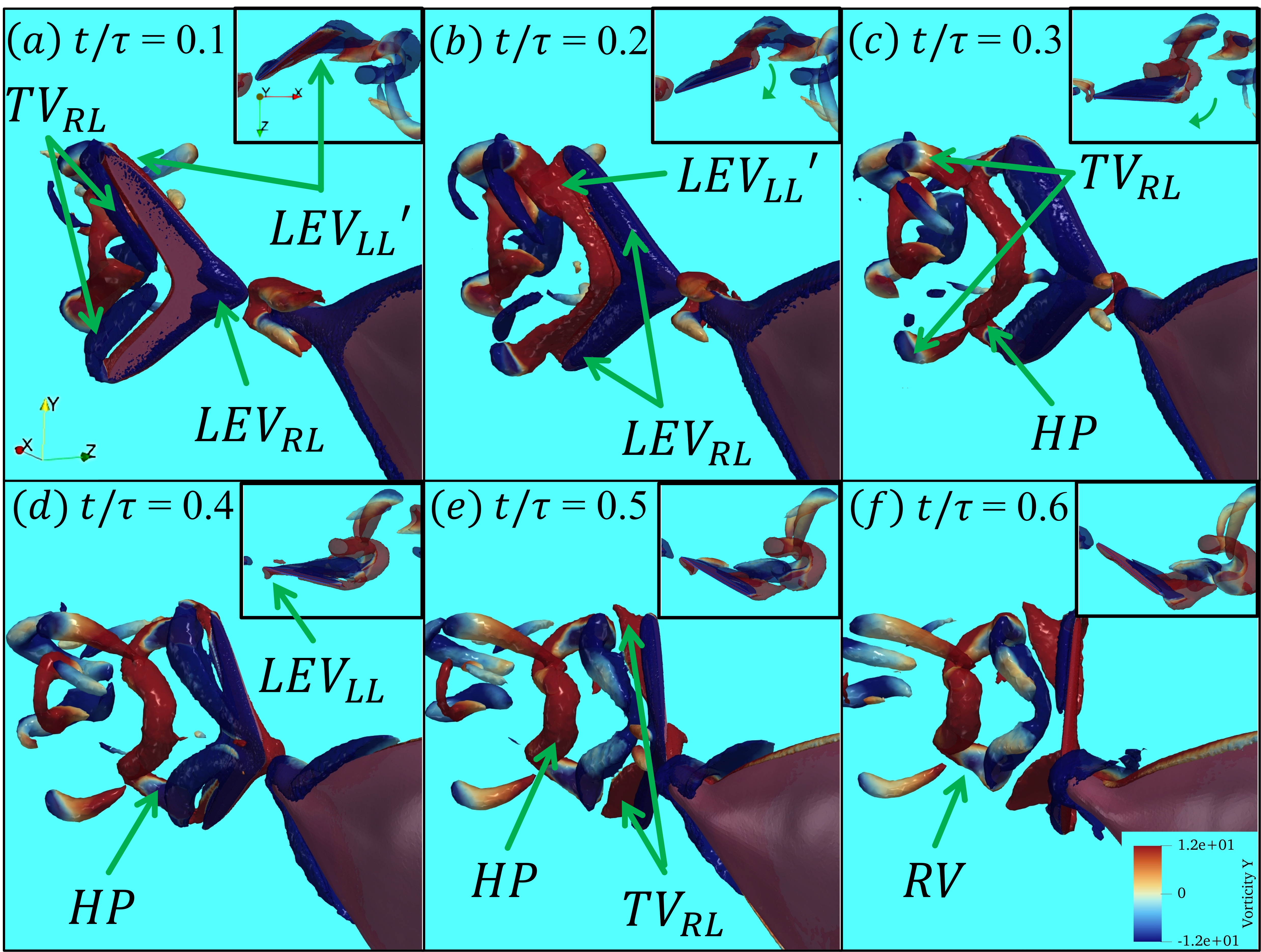}
    \caption{Close up views for vortices generated and sheded in a half oscillation cycle from the caudal fin for case $5$, where \edt{the vortices are visualized by iso-surface of} $Q = 40$ colored by \edt{$\omega_y$}.}
    \label{fig:case5VortexDynamics}
\end{figure}

Figure~\ref{fig:case5VortexDynamics}a–$f$ presents the vorticity contours for case~$5$ over half of an oscillation cycle, during which the caudal fin pitches from the right ventral side toward the left ventral side. The pitching \edt{kinematics here} can be divided into two distinct \edt{stages}. \edt{Stage}~1 ($t/\tau = 0.1$–$0.3$) corresponds to the fin pitching from the right ventral side toward the left ventral side as it approaches the neutral axis. \edt{Stage}~2 ($t/\tau = 0.4$–$0.6$) begins once the fin crosses the neutral axis and continues as it gradually pitches toward the peak on the left ventral side. In Fig.~\ref{fig:case5VortexDynamics}, each snapshot includes a dorsal view inset in the upper right corner, which assists in tracking the instantaneous orientation of the fin and examining the evolution of vortical structures on the left ventral side. The vortices in the isometric view are visualized using a $Q$-criterion threshold of $12$, while the dorsal view employs both $Q = 12$ and a translucent $Q = 8$ to highlight coherent structures.


Figure~\ref{fig:case5VortexDynamics}a captures the instant immediately after the caudal fin begins its pitching stroke from the right ventral side towards the left ventral side. At this stage, tip vortices ($\mbox{TV}_{\mbox{LL}}$) forms at both the upper and lower lobes, and a leading-edge vortex ($\mbox{LEV}_{\mbox{RL}}$) develops on the right ventral side. Meanwhile, the leading-edge vortex on the right ventral side from the previous cycle ($\mbox{LEV}_{\mbox{LL}}$) remains attached to the fin surface. Advancing to the next instance in Fig.~\ref{fig:case5VortexDynamics}b, the leading-edge vortex ($\mbox{LEV}_{\mbox{RL}}$) continues to strengthen on the right ventral side and the tip vortex ($\mbox{TV}_{\mbox{RL}}$) on the same side sheds into the wake. In Fig.~\ref{fig:case5VortexDynamics}c, the upper and lower tip \edt{vortices} sheds into the wake. The convected $\mbox{LEV}_{\mbox{LL}}$', rolls downstream along the trailing-edge span before \edt{its detachment from the fin}, stretches into a structure that mirrors the geometry of the trailing-edge span and remains connected to $\mbox{LEV}_{\mbox{RL}}$ through the upper and lower lobes. This combined structure closely resembles a hairpin vortex ($\mbox{HP}$). As noted earlier in Fig.~\ref{fig:waveComparison}, the caudal fin pitches away from its peak amplitude rapidly, promoting the formation of this hairpin vortex. The \edt{structure} of the resulting $\mbox{HP}$ is influenced by the geometry of the fin, and in this case, it adopts the characteristic curvature of the trailing-edge span.

Moving into Stage~2, the pitching of the caudal fin slows as it crosses the neutral axis and approaches the end of the half-stroke, where it reaches its maximum heaving amplitude. In Figs.~\ref{fig:case5VortexDynamics}d–-\edt{\ref{fig:case5VortexDynamics}}f, a portion of the right ventral-side vortex ($\mbox{LEV}_{\mbox{RL}}$) begins to detach from the surface while remaining connected to the hairpin vortex ($\mbox{HP}$) through the upper and lower lobes. The dorsal-view insets show the emergence of a new leading-edge vortex ($\mbox{LEV}_{\mbox{LL}}$) from the left ventral side as the angle-of-attack increases. The gradual pitching in this phase prevents a complete separation of $\mbox{LEV}_{\mbox{RL}}$, \edt{which} is an important feature that helps the swimmer improve its thrust \edt{significantly \cite{borazjani2013fish}}. In Fig.~\ref{fig:case5VortexDynamics}e, $\mbox{LEV}_{\mbox{RL}}$ \edt{gets stronger in terms of its growing size} while remaining attached to the right ventral side of the fin. \edt{Simultaneously}, the previously formed $\mbox{HP}$, which \edt{previously remains} connected to $\mbox{LEV}_{\mbox{RL}}$, fully detaches from the fin and stretches into a distorted ring-like coherent structure. As the angle-of-attack continues to \edt{increase}, new tip vortices emerge from the upper and lower lobes on the left ventral side. In Fig.~\ref{fig:case5VortexDynamics}f, the distorted structure observed at the earlier instant evolves into a more complete ring vortex, once again adopting a geometry that closely follows the shape of the trailing-edge span of the caudal fin.

\begin{figure}[ht!]
    \centering
    \includegraphics[width=1\linewidth]{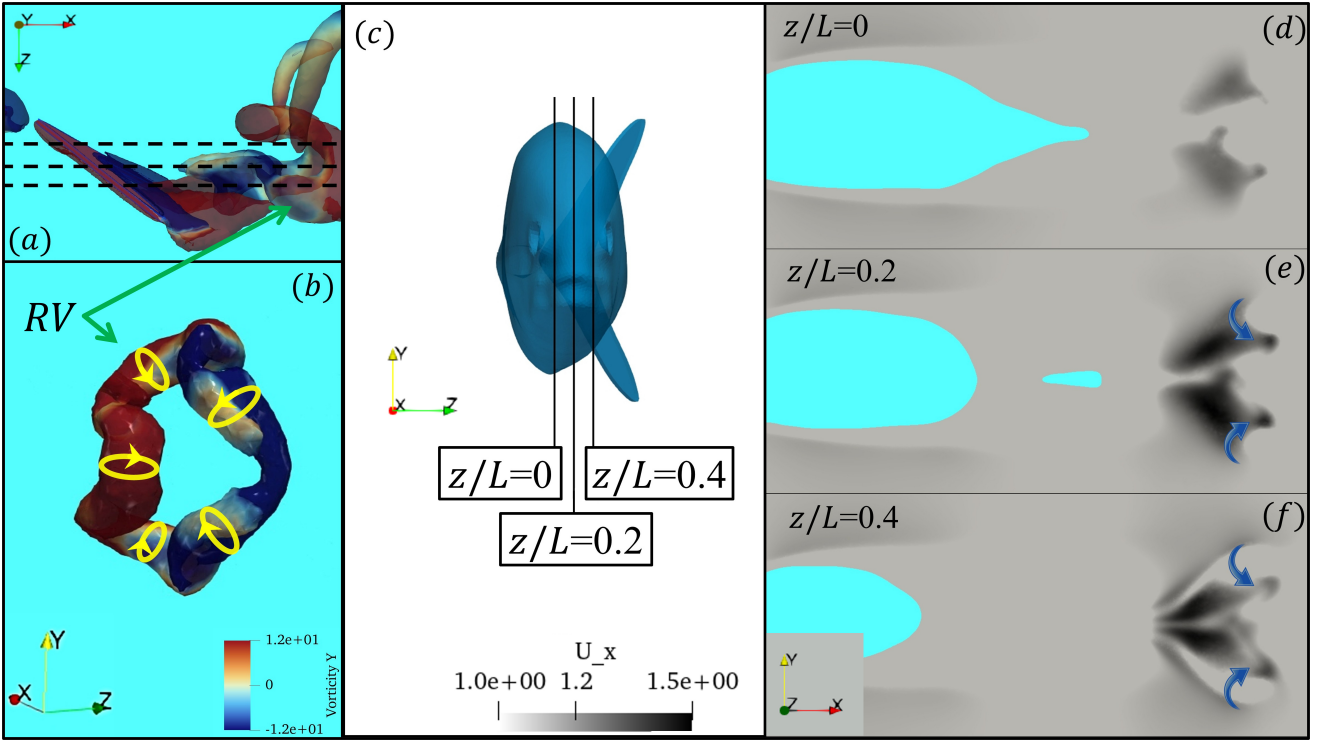}
    \caption{\edt{Structures of} the ring vortex generated in case~5 at $t/\tau = 0.6$, shown from (a) the dorsal view and (b) the isometric view, \edt{whereas the plot in} (c) indicates the locations of the iso-surface slices along the $z$-axis from the anterior view of the swimmer, corresponding to images (d)–-(f), which display the streamwise velocity contours extracted from each slice.}
    \label{fig:CASE5_VORTEXtHRUST}
\end{figure}

We can examine the influence of the ring vortex observed in Fig.~\ref{fig:case5VortexDynamics}$f$ using the instantaneous streamwise velocity distribution. \edt{Figure}~\ref{fig:CASE5_VORTEXtHRUST}a presents the dorsal view of the caudal fin, showing the geometry of the ring vortex in the $X$–$Z$ plane along with the three dashed lines that indicate the locations of the slices used for the illustration for the velocity\edt{-based} analysis. The anterior view in Fig.~\ref{fig:CASE5_VORTEXtHRUST}b further clarifies the spatial placement of these slices relative to the fin. Velocity contours extracted from the $Y$–$Z$ planes at $z/L = 0$, $0.2$, and $0.4$, shown in Figs.~\ref{fig:CASE5_VORTEXtHRUST}d,~\edt{\ref{fig:CASE5_VORTEXtHRUST}}e, and~\edt{\ref{fig:CASE5_VORTEXtHRUST}}f, respectively, \edt{show} the streamwise velocity in the range $U_x = 1$ to $1.5$ (normalized as $U_x = \dot{x}/\edt{U_\infty}$). Figure~\ref{fig:CASE5_VORTEXtHRUST}\edt{b} isolates the ring vortex, where the red color contour represents the anti-clockwise rotation of the vortex about $Y$-axis and the blue color shows the clockwise rotation of the vortex.The inward folding ring implies an increase in the spanwise velocity as the vortex \edt{is} shed downstream. Examining Figs.~\ref{fig:CASE5_VORTEXtHRUST}d–\ref{fig:CASE5_VORTEXtHRUST}f, a clear jet-like effect is observed in the wake at the instant where the ring vortex is formed. In \edt{Fig.}~\ref{fig:CASE5_VORTEXtHRUST}(e), the\edt{peak of the} streamwise velocity is most pronounced, corresponding to the center region between the upper and lower lobes of the ring vortex, where its inward-directed rotation contributes to the streamwise momentum.

\begin{figure}[ht!]
    \centering
    \includegraphics[width=1\linewidth]{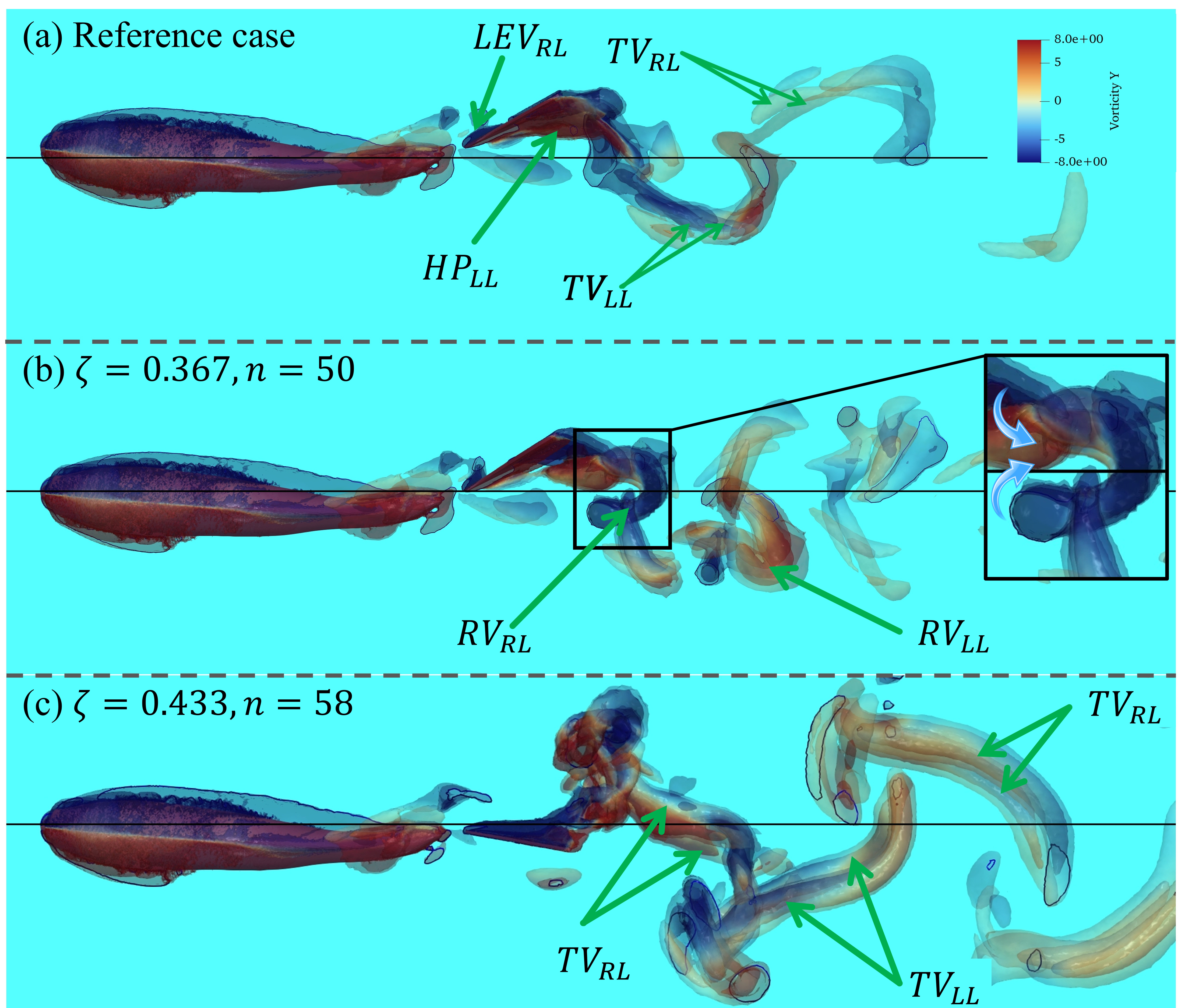}
    \caption{\edt{Vortex visualization around the swimmer using iso-surfaces at $Q=40$, and $Q=8$, colored by $\omega_y$} at the end of an oscillation cycle for (a) \edt{the} case with an actively pitching tail, (b) case $5$, and (c) case $9$. }
    \label{fig:vorticityComparison}
\end{figure}

\edt{To better understand the thrust-dominated behavior described above, we examine} the dorsal-view vortex dynamics at the end of a full oscillation cycle ($t/\tau = 0.9$) for the actively pitching case, case~5, and case~9, \edt{as} shown in Figs.~\ref{fig:vorticityComparison}a–-\ref{fig:vorticityComparison}c, respectively. These snapshots allow the \edt{fin's} posture to be visually correlated with the vortical structures present in the wake. In Fig.~\ref{fig:vorticityComparison}a, the actively pitching \edt{tail} sheds two tip vortices from the right ventral side ($\mbox{TV}_{\mbox{RL}}$) and two from the left ventral side ($\mbox{TV}_{\mbox{LL}}$) over one cycle. \edt{It is observed that} the $\mbox{LEV}_{\mbox{RL}}$ \edt{is formed} on the caudal fin, and a hairpin vortex emerges from the trailing-edge span. This hairpin structure reflects the geometry of the fin, with distinct upper and lower lobes, which then stretches downstream to form the tip vortices observed in the wake. In Fig.~\ref{fig:vorticityComparison}b, case~5 exhibits the ring vortices previously analyzed, with their inward rotational direction clearly visible and contributing to the strengthening of the downstream jet associated with \edt{production of thrust}. \edt{On the contrary}, Fig.~\ref{fig:vorticityComparison}c for case~9 reveals several notable features. First, the pronounced $\mbox{PAS}$ is evident from the larger displacement of the caudal fin relative to both the active \edt{pitching tail} and case~5. Second, the wake pattern contains two right ventral-side and two left ventral-side tip vortices \edt{that are} similar to those \edt{from the actively pitching tail}, yet the spatial distribution differs. The vortices in case~9 display significantly greater lateral spreading, indicating a broader wake footprint compared to the actively pitching fin.

\begin{figure}[ht!]
    \centering
    \includegraphics[width=1\linewidth]{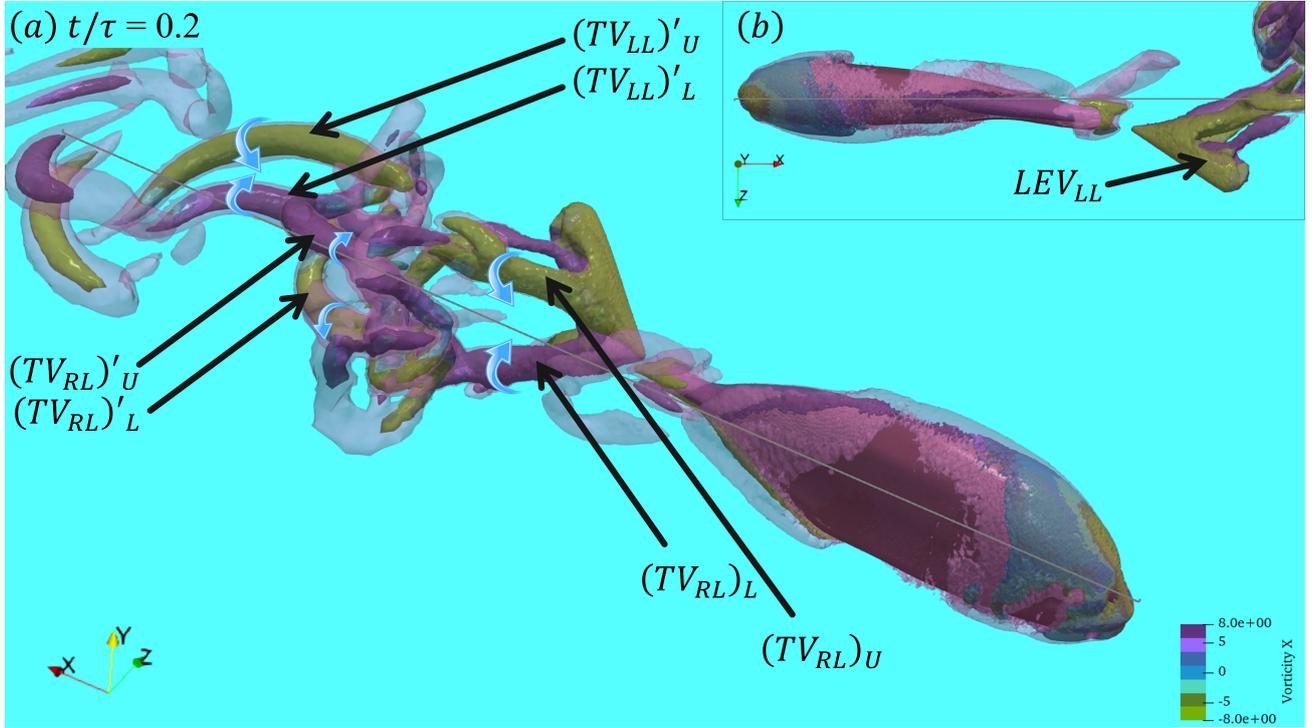}
    \caption{{Vortex visualization around the swimmer using iso-surfaces at $Q=40$, and $Q=8$, colored by $\omega_x$} (a) iso-matric view, and (b) the dorsal view at the end of an oscillation cycle with the iso-surfaces at $Q=40$, and translucent $Q=8$}
    \label{fig:case9VortexDynamics}
\end{figure}

{It requires further examination to understand why case~9 is not hydrodynamically advantageous for the swimmer. In order to perform this analysis, Fig.}~\ref{fig:case9VortexDynamics}a presents the isometric view of the swimmer, while Fig.~\ref{fig:case9VortexDynamics}b shows the corresponding dorsal view. The coherent structures in the wake display strong extension in the streamwise direction, motivating the use of \edt{$\omega_x$} to color the $\mbox{3D}$ vortices \edt{for characterizing them}. In Fig.~\ref{fig:case9VortexDynamics}a, the tip vortices shed from the caudal fin are evident\edt{. However}, unlike cases where the vortices originate at the trailing-edge span, these structures \edt{are detached} directly from \edt{$\mbox{LEV}_{\mbox{LL}}$} as they \edt{are formed and separated} from the surface. The previously shed vortices are labeled with a $'$ as $(\mbox{TV}_{\mbox{RL}})'_U$ and $(\mbox{TV}_{\mbox{RL}})'_L$ for the upper and lower right ventral-side vortices, and $(\mbox{TV}_{\mbox{LL}})'_U$ and $(\mbox{TV}_{\mbox{LL}})'_L$ for their left ventral-side counterparts.

The newly formed right ventral-side vortices, $(\mbox{TV}_{\mbox{RL}})_U$ and $(\mbox{TV}_{\mbox{RL}})_L$, appear as the angle-of-attack increases, indicating a premature separation of the right ventral-side vortex. Such early detachment \edt{has a drastic effect on thrust production. An} examination of their rotational direction shows inward rotation, similar to what \edt{is} observed in case~5 (see Fig.~\ref{fig:case9VortexDynamics}). However, this inward rotation in case~9 induces a more pronounced spanwise spreading of the vortex, \edt{contrary} to case~5 where the ring vortex \edt{strengthens} the streamwise jet. Here, the dominant lateral influence suggests that the vortex dynamics favor wake broadening rather than \edt{the} axial velocity.

\begin{figure}[ht!]
    \centering
    \includegraphics[width=1\linewidth]{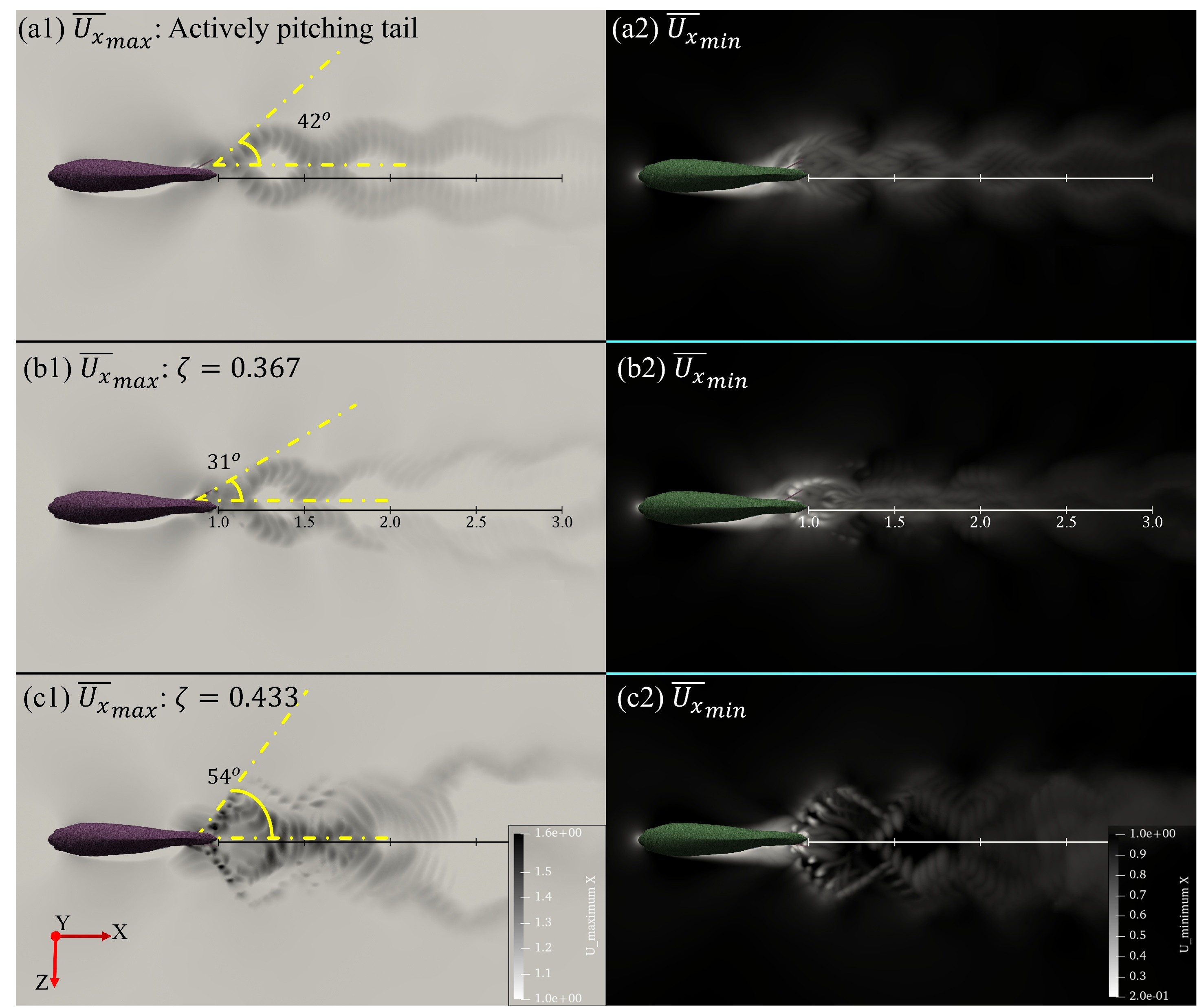}
    \caption{\edt{Contours} for the cycle-averaged value for the case with ($a1$ \& $a2$) an actively pitching tail, ($b1$ \& $b2$) case $5$, and ($c1$ \& $c2$) case $9$ referring to $\overline{U_x}_{max}$, $\overline{U_x}_{min}$, respectively.}
    \label{fig:cycleAverageMinMax}
\end{figure}

\edt{Now,} the contours of the maximum and minimum cycle-averaged streamwise velocity for the actively pitching case, case~5, and case~9 \edt{are shown} in Figs.~\ref{fig:cycleAverageMinMax}a(1–2), b(1–2), and c(1–2), respectively. \edt{Figures}.~\ref{fig:cycleAverageMinMax}$a1$, \ref{fig:cycleAverageMinMax}$b1$, and \ref{fig:cycleAverageMinMax}$c1$ correspond to the maximum cycle-averaged streamwise velocity, while Figs.~\ref{fig:cycleAverageMinMax}$a2$, \ref{fig:cycleAverageMinMax}$b2$, and \ref{fig:cycleAverageMinMax}$c2$ present the corresponding minimum values. For the actively pitching fin, \edt{the distribution of} $\overline{U_x}_{\text{max}}$ exhibits a well-defined diverging jet, forming an inclination angle of approximately $42^\circ$ with the centerline. The contours gradually weaken downstream, reflecting the decay of streamwise momentum. \edt{On the other hand}, the $\overline{U_x}_{\text{min}}$ field in Fig.~\ref{fig:cycleAverageMinMax}($a2$) displays a narrow wake footprint that highlights regions of \edt{a decreased} velocity relative to the freestream \edt{one}. This low-velocity region dissipates rapidly, indicating that the associated vortical shedding exerts a neutral influence and do not contribute significantly to thrust or drag.

In Fig.~\ref{fig:cycleAverageMinMax}($b1$), case~5 shows a jet inclination angle of approximately $31^\circ$, smaller than that of the actively pitching fin but with a wake pattern that closely resembles it. The $\overline{U_x}_{\text{min}}$ field in Fig.~\ref{fig:cycleAverageMinMax}($b2$) reveals an even narrower and weaker region of reduced velocity, consistent with the strong thrust observed for this configuration. \edt{Nevertheless}, case~9 exhibits the largest jet inclination angle, reaching nearly $54^\circ$ in Fig.~\ref{fig:cycleAverageMinMax}($c1$). Although a strong streamwise jet persists up to roughly $2.0L$, the wake bifurcates beyond this distance. Examination of Fig.~\ref{fig:cycleAverageMinMax}($c2$) \edt{highlights} $\overline{U_x}_{\text{min}} < U_\infty$ near the centreline, producing a low-momentum region associated with \edt{a} reduced thrust. This behavior aligns with the broader, more diffused wake generated by the pronounced $\mbox{PAS}$ in case~9. As also reported by Dong et al. \cite{Dong2006Wake}, the larger jet inclination angle correlates to the drop in thrust.
\section{Conclusions}
This study demonstrates that nonlinear torsional stiffness at the peduncle can passively regulate caudal fin kinematics and deliver hydrodynamic performance comparable to that of an actively pitching tail. The displacement-dependent stiffness enables large pitching amplitudes while stabilizing the fin near its extreme positions, producing a natural recoil mechanism that governs both amplitude and phase synchronization. When the stiffness parameters are tuned to achieve synchronization with the undulation of the biody, the caudal fin generates coherent hairpin and ring vortices that leads to streamwise momentum and promote a thrust-dominant wake. In contrast, larger phase mismatches shift the wake structure toward a drag-dominated regime characterized by pronounced lateral vortex spreading. These findings demonstrate that nonlinear stiffness at the peduncle functions not only as a structural element but also as a hydrodynamically advantageous mechanism that enables passive synchronization through fluid–structure interaction. The results provide new physical insight into carangiform swimming strategies and establish design guidelines for nature-inspired robotic platforms capable of achieving propulsion without active caudal fin pitching.
\section{Acknowledgment}

MSU Khalid acknowledges funding support from the Natural Sciences and Engineering Research Council of Canada (NSERC) through the Discovery and Alliance International grant programs for this work. A. Tarokh also thanks NSERC for their support though the Discovery grant. The simulations reported in this work were performed on the supercomputing clusters administered and managed by the Digital Research Alliance of Canada. 
\bibliography{references}

\end{document}